\documentclass[aps,prd,showpacs,
twocolumn
,superscriptaddress]{revtex4}
\usepackage{graphicx}% Include figure files
\usepackage{dcolumn}% Align table columns on decimal point
\usepackage{bm}% bold math
\usepackage{color}
\usepackage[normalem]{ulem} % \sout{old text} for strikeout
\usepackage[dvipsnames]{xcolor} % For blue in-text comments
\usepackage{hyperref}
\hypersetup{
  colorlinks=true,        % false: boxed links; true: colored links
  linkcolor=blue,         % color of internal links
  citecolor=cyan,         % color of links to bibliography
}
\usepackage{mathrsfs}
\usepackage[utf8]{inputenc}
\usepackage{mathtools}
\usepackage{doi}
\usepackage{amsmath}
\usepackage{amssymb}

\begin{document}

\title{Dynamics of test particles around renormalization group improved Schwarzschild black holes}

\author{Javlon Rayimbaev}
\email{javlon@astrin.uz}
\affiliation{Ulugh Beg Astronomical Institute, Astronomicheskaya 33, Tashkent 100052, Uzbekistan}
\affiliation{National University of Uzbekistan, Tashkent 100174, Uzbekistan}
\affiliation{Institute of Nuclear Physics, Ulugbek 1, Tashkent 100214, Uzbekistan}

\author{Ahmadjon~Abdujabbarov}
\email{ahmadjon@astrin.uz}

\affiliation{Ulugh Beg Astronomical Institute, Astronomicheskaya 33, Tashkent 100052, Uzbekistan}
\affiliation{National University of Uzbekistan, Tashkent 100174, Uzbekistan}
\affiliation{Institute of Nuclear Physics, Ulugbek 1, Tashkent 100214, Uzbekistan}
\affiliation{Shanghai Astronomical Observatory, 80 Nandan Road, Shanghai 200030, P. R. China}
\affiliation{Tashkent Institute of Irrigation and Agricultural Mechanization Engineers, Kori Niyoziy, 39, Tashkent 100000, Uzbekistan}

\author{Mubasher~Jamil}
\email{mjamil@zjut.edu.cn (corresponding author)}
\affiliation{Institute for Theoretical Physics and Cosmology, Zhejiang University of Technology, Hangzhou 310023 China}
\affiliation{School of Natural Sciences, National University of Sciences and Technology, Islamabad, 44000, Pakistan}
 \affiliation{Canadian Quantum Research Center, 204-3002, 32 Ave, Vernon, BC, V1T 2L7, Canada}
\author{Bobomurat~Ahmedov}
\email{ahmedov@astrin.uz}
\affiliation{Ulugh Beg Astronomical Institute, Astronomicheskaya 33, Tashkent 100052, Uzbekistan}
\affiliation{National University of Uzbekistan, Tashkent 100174, Uzbekistan}
\affiliation{Tashkent Institute of Irrigation and Agricultural Mechanization Engineers, Kori Niyoziy, 39, Tashkent 100000, Uzbekistan}

\author{Wenbiao Han}
\email{wbhan@shao.ac.cn}
\affiliation{Shanghai Astronomical Observatory, 80 Nandan Road, Shanghai 200030, P. R. China}

\date{\today}

\begin{abstract}

In this paper we have investigated the dynamics of neutral, electrically charged and magnetized particles around renormalized group improved (RGI) Schwarzschild black hole in the presence of external asymptotically uniform magnetic field. We have analyzed the spacetime structure around RGI black hole by investigating Ricci, the square of Ricci tensor and Kretschmann curvature scalars and shown that only in the case when the parameter $\gamma=0$ the curvature becomes infinite at the center of the black hole, while for non-zero values of $\gamma$ parameter the black hole curvature reflects the properties of regular black hole. Analyzing the innermost stable circular orbits of test neutral particles around RGI black hole and comparing with the results for rotating Kerr black hole we have shown that RGI black hole parameters can mimic the rotation parameter of Kerr black hole upto $a/M \lesssim 0.31$ providing the same ISCO radius. Since according to the astronomical observations of the accretion disks confirm that the astrophysical black holes are rapidly rotating with the spin parameter upto $a/M \sim 0.99$ one may conclude that the effects of parameters of RGI Schwarzschild black hole on the circular orbits of the neutral particles can not mimic the Kerr black hole.  Then the Hamilton-Jacobi equation has been used to analyze the charged and magnetized particles motion near the RGI black hole in the presence of the strong interaction between external asymptotically uniform magnetic (electromagnetic) field and magnetized (electrically charged) particle. We have shown that RGI black hole parameters quantitatively change the dynamics of the charged and magnetized particles, in particular ISCO radius of the particles decreases with increasing the parameter $\lambda$, while the increase of the parameter $\gamma$ causes to increase of it. The stability analysis of the circular orbits of the magnetized particles has shown that the population of magnetars with the strong surface magnetic field upto $10^{15}$ Gauss is excluded from the very close environment of a SMBH due to destructive properties of the magnetic field.   

\end{abstract}

\maketitle

\section{Introduction}

The direct and indirect evidences about the presence of BHs (BHs) within the stellar and galactic environments in the mass range ($\sim 3-10^{10} M_\odot$) have been obtained by numerous astronomical observations see e.g. \cite{Bambi17c}.  The astrophysical BHs which are typically found in the X-ray binaries have masses $3\sim 20 M_\odot$, while more massive BHs of order 30 $M_\odot$ are found via gravitational wave detectors before the merger of stellar mass BHs in close BH binaries forming up to $\sim 60 M_{\odot}$ BH at the end state. Supermassive BHs at the center of galaxies observed as active galactic nuclei (AGN) with extremely high luminosity at $\sim 10^{45}\rm erg \cdot s^{-1}$  have masses in the range $10^5-10^{10} M_\odot$ \cite{Bambi17e}. In the astrophysical setting, the astrophysical BHs can be found in the relative states either dormant or active. In the former case, the geometry and nearby environment of the BH can be studied, for instance via its shadow and gravitational lensing while in the later case, the  supermassive BH is found along with X-ray (and/or ultraviolet) emitting accretion disks and jets \cite{Falcke02}. In the recent years, the Event Horizon Telescope (EHT), GRAVITY and LIGO-VIRGO Collaborations have provided  first observational data and new information about the formation and the nearest regions of BHs in strong field regime \cite{EHT19a,LIGO16}, thereby confirming the predictions of general relativity and also providing the powerful tools to get necessary constraints on modified gravity theories. 

A BH is essentially composed of two parts namely, the event horizon and the physical curvature singularity, which is  fundamentally unobservable by any means of observations. The hypothetical event horizon of the BH can be directly and indirectly observed the observation of dynamics of photons and matter in very close environment of BH. Therefor, BHs serve truly as astrophysical laboratories for testing various gravity theories in the strong gravity regime. In particular, using the motion of test particles and following their orbits near the BH, the geometrical structure of spacetime can be determined \cite{Jawad16,Hussain15,Jamil15,Hussain17,Babar16,Banados09,Majeed17,Zakria15,Brevik19}. If the BH is surrounded by a test uniform magnetic field, than the trajectories of charged particles are affected as well, resulting in the chaotic behavior \cite{Chen16,Hashimoto17,Dalui19,Han08,Moura00}. 
The first study of the electromagnetic field structure around a BH has been carried out by Wald~\cite{Wald74}.  
The magnetic field is assumed to be weak and the test particle is assumed to have negligible mass compared to BH mass, such that back-reaction effects are ignored in the analysis. As a result, a number of interesting phenomena are studied in the literature including center of mass energy produced via particle collisions, generation of chaos in particle motion, geodesic precession frequencies, quasi-periodic oscillations and weak deflection angle, to name a few \cite{Azreg20,Liu20,Jusufi20}.

In the presence of external magnetic field around BH one may  consider the magnetized particle motion. The pioneer works on this subjects have been carried out in Refs.~\cite{deFelice,defelice2004}. 
The magnetized particle motion around non-Schwarzschild BH in the presence of a magnetic field has been studied in~\cite{Rayimbaev16}. Acceleration of magnetized particle around a rotating BH in quintessence has been considered in~\cite{Oteev16} . Magnetized particles collision has been also explored in Refs.~\cite{Toshmatov15d,Abdujabbarov14,Rahimov11a,Rahimov11} for the different gravity models. Our recent works have been devoted to study of the magnetized particle motion in conformal gravity, 4D Einstein-Gauss-Bonnet gravity and modified gravity models~\cite{Haydarov20,Haydarov2020EPJC,Rayimbaev4DEGB2020,Vrba2020PhRvD}.
The properties of the electromagnetic field around BH in the presence of external asymptotically uniform and dipolar magnetic field have been explored in~\cite{Kolos17,Kovar10,Kovar14,Aliev89,Aliev02, Aliev86,Frolov11,Frolov12,Shaymatov18,Stuchlik14a, Abdujabbarov10, Abdujabbarov11a,Abdujabbarov11,Abdujabbarov08,Karas12a,Stuchlik16,Rayimbaev20,Turimov18b,Turimov17,Rayimbaev15,Rayimbaev19,Narzilloev2020CG,Rayimbaev2020PRD,Nathanail2017MNRAS,MorozovaV2014PhRvD}.

A fundamental problem in general relativity is that it predicts the existence of singularities with infinite curvature in spacetime which breaks down in their vicinity. In particular, any perturbative approach to quantify general relativity leads to divergences and non-renormalizability. However, imposing suitable cut-offs at the high energy limit in quantum gravity, the quantization problem can be circumvented \cite{Mukhanov07,Gambini11,Oriti09}. In particular, the quantum gravitational effects in BHs has a long history with profound results including the Hawking evaporation process, entanglement and information loss paradox. The quantum gravity effects followed from the loop quantum gravity and the renormalization-group-improvement (RGI) approach, resolve the issue of BH singularities i.e. the formation of singularity during the gravitational collapse of a dying star is either delayed or completely avoided \cite{Bonanno00,Lu19}. In the Hawking evaporation process, it is likely that the BH is completely evaporated and the curvature singularity gets naked.  However, the RGI approach proposes that the evaporation process terminates at the stage when critical mass is reached thus no singularity appears at the end state. This theory also replaces the Schwarzschild singularity at $r=0$ with a de Sitter core. In literature, the RGI Schwarzschild BH has been studied in the wider context including thermodynamics of horizons, gravitational lensing and the accretion dynamics \cite{Lu19,Yang15}.
 
In this paper, we consider neutral, electrically charged and magnetized particles motion around BH within renormalization-group-improvement approach. The paper is organized as follows: We start with the review of the solution of BH in RGI model and explore the spacetime curvature properties in Sect.~\ref{chapter1}. We study the test particle dynamics around BH in RGI model in Sect~\ref{testpartmotion}.  We study the charged and magnetized particles dynamics in Sects~\ref{chargedpartmotion} and \ref{magnetpartmotion}, respectively. We conclude our results in Sect.~\ref{conclusion}. 

The spacelike signature $(-,+,+,+)$  is selected 
for the spacetime and the geometrized system is used where $G = c= h = 1$ (However, for an astrophysical application the speed of light and Newtonian gravitational constant are written explicitly in our expressions). The Latin indices run from $1$ to $3$ and the Greek ones from $0$ to $3$.

\section{The spacetime properties \label{chapter1}}

The quantum effects in the geometry of Schwarzschild black hole are incorporated using the techniques of renormalization group. Here the Newton's gravitational constant is considered to be varying with a length scale. A consequence of this modification is that the resulting geometry is free from spacetime singularities, however it does admit none, one or two horizons. Another ramification is that the black hole evaporation terminates when the black hole mass reaches to a critical mass, a cold remnant with an AdS geometry~\cite{Bonanno00}. The central notion behind renormalization group approach is that the effective description of quantum field theory in a curved background must be UV complete and non-perturbatively renormalizable using asymptotic safety \cite{Weinberg1978}. In literature, few new black hole solutions have been derived in the asymptotically safe gravity theory and analyzed \cite{Cai2010JCAP,Haroon2018EPJC,Farooq_2020,Eichhorn2017PhRvD,Platania2019EPJC}.
We now analyze the spacetime metric around RGI Schwarzschild black hole. The line element of the spacetime around the static BH with mass $M$  can be described as~\cite{Bonanno00}
\begin{eqnarray}\label{metric}
ds^2=-f(r)dt^2+\frac{1}{f(r)}dr^2+r^2 (d\theta^2 +\sin^2\theta d\phi^2)\ ,
\end{eqnarray}
with  
\begin{equation} \label{metfunct}
f(r)=1-\frac{2M}{r}\Big(1+ \frac{\Omega M^2}{r^2}+\frac{\gamma\Omega M^3}{r^3}  \Big)^{-1} , 
\end{equation}
where $\gamma$ and $\Omega$ are the new spacetime parameters motivated by the non-perturbative renormalization group theory and characterize quantum corrections. The number of horizons depend on the number of roots of the equation $f(r)=0$, which result in none, single or double horizons. The discriminant of the cubic equation is $\Delta=-M^6\Omega(\Omega-\Omega_+)(\Omega-\Omega_-),$ where
$$\Omega_\pm=-\frac{27}{8}\gamma^2-\frac{9}{2}\gamma+\frac{1}{2}\pm\frac{1}{8}\sqrt{(\gamma+2)(9\gamma+2)^3}.$$
Here the cubic equation has none or double positive roots provided $\Omega>0$ and $\gamma>0$. For $\Delta>0$, we require $\Omega\leq\Omega_+$ and $\Omega_- <0$. Henceforth a new dimensionless parameter $\lambda$ was introduced as follows \cite{Lu19}: $\Omega=\lambda\Omega_+$, where $0<\lambda\leq1$. For $0<\lambda<1$, two horizons exist; $\lambda=1$ yields a single horizon (the extremal case) while $\lambda>1$ gives no horizon. It is important to note that both $\lambda$ and $\gamma$ are free parameters in the theory and can be constrained only through some means of astrophysical data. Recently, Lu and Xie investigated the weak and strong gravitational lensing by the RGI Schwarzschild black hole by taking the parameters of the supermassive black holes at the center of M87 and Milky Way galaxy \cite{Lu19}. They concluded that the data concerning the shadow of M87 central black hole provide the following bounds on the parameters as $0.2\leq\lambda\leq10$ and $0.02\leq\gamma\leq0.22$. Given these numerical values, one can also constrain the parameter $\Omega$ as $0.165\leq\Omega\leq9.804$. Although these parameters need to be constrained tightly by considering alternative astrophysical sources such as gravitational waves generated by the black hole mergers, however the current constraints allow the possibility of one or two horizons of the RGI black hole and consequently should be investigated. 

 \subsection{Scalar invariants}\label{RRKinvariants}
 
In fact the detailed analysis of curvature invariants such as the Ricci scalar, square of the Ricci tensor and the Kretschmann scalar may give deep understanding of the main properties of the spacetime. By that reason, in this section, we investigate the curvature invariants of thespacetime  metric (\ref{metric}).

{\bf The Ricci scalar. }
First, we explore the effects of the spacetime parameters $\lambda$ and $\gamma$ on the value of Ricci scalar. The expression for the Ricci scalar can be easily derived in the following form 
\begin{eqnarray}\label{R}
 R&=&g^{\mu \nu}R_{\mu \nu}=\frac{4 \lambda  M^3 \Omega_+  }{\left[\lambda  M^2 \Omega_+  (\gamma  M+r)+r^3\right]^3}\Big[\lambda  M^2 \Omega_+ \nonumber\\ 
 && (6 \gamma^2 M^2+8 \gamma  M r+3 r^2)-r^3 (3 \gamma  M+r)\Big]\ .
\end{eqnarray}

One can see from Eq(\ref{R}) that in the Schwarzschild limit when $\lambda=0$ the spactime becomes  the Ricci flat one. Moreover, the Ricci scalar is also equal to zero at the limit when  $r \to r_{\rm h}$, it implies that the spacetime is Ricci flat out of event horizon and, consequently one may consider the spacetime as an asymptotically Ricci flat one. Now we will analyze the Ricci scalar of the spacetime (\ref{metric}) in the different limits, namely at $\gamma=0$ and $\gamma \to \infty$:
\begin{eqnarray}\label{R1}
 \lim_{\gamma \to 0}R&=&\frac{4 \lambda  \left(3 \lambda-r^2 \right)}{r \left(\lambda +r^2\right)^3}\ ,
 \\
 \lim_{\gamma \to \infty}R&=&\frac{10368 \lambda  \left(64 \lambda-27 r^3 \right)}{\left(32 \lambda +27 r^3\right)^3}\ .
\end{eqnarray}
It is easy to see that there is singularity at the center of the spacetime of the RGI Schwarzschild  BH when $\gamma=0$. However, for the nonvanishing $\gamma\neq0$ (even $\gamma \to \infty$) the spacetime has finite curvature at $r=0$ being a regular BH and we have:
\begin{eqnarray}\label{R2}
 &&\lim_{r \to 0}R=\frac{3 \left[\sqrt{\gamma +2} (9 \gamma +2)^{3/2}+9 \gamma  (3 \gamma +4)-4\right]}{8 \gamma ^2 \lambda  M^2} , \ \ \\
&& \lim_{\gamma \to \infty} \left( \lim_{r \to 0}R\right)=\frac{81}{4 \lambda  M^2}\ .
 \end{eqnarray} 

{\bf The square of Ricci tensor. }
Now  we explore the square of Ricci tensor of spacetime (\ref{metric}) which can be found in the following form 
 \begin{eqnarray}\label{RR}
 R_{\mu \nu} R^{\mu \nu}&=&\frac{8 \lambda ^2 M^6 \Omega_+^2 }{\left[\lambda  M^2 \Omega_+  (\gamma  M+r)+r^3\right]^6} 
\nonumber \\\nonumber
&\times & \Big[r^6 (45 \gamma ^2 M^2+48 \gamma  M r+13 r^2)
    \\\nonumber
  &+ & 2 \lambda  M^2 r^3 \Omega_+  (r^3 +\gamma  M r^2-6 \gamma ^2 M^2 r
  \\ \nonumber
 &-& 9 \gamma ^3 M^3)+\lambda ^2 M^4 \Omega_+^2  (18 \gamma ^4 M^4+48 \gamma ^3 M^3 r
 \\ 
 &+&52 \gamma ^2 M^2 r^2+26 \gamma  M r^3+5 r^4)\Big]\ .
\end{eqnarray}
One may easily see that in the case when $\lambda=0$ the square of the Ricci tensor also tends to zero as well as Ricci scalar. In the other limits it has the following form: 
\begin{eqnarray}\label{RR1}
 \lim_{\gamma \to 0}R_{\mu \nu} R^{\mu \nu}&=&\frac{8 \lambda ^2 \left(5 \lambda ^2+13 r^4+2 \lambda  r^2\right)}{r^2 \left(\lambda +r^2\right)^6}\ ,
 \\\nonumber
\lim_{\gamma \to \infty}R_{\mu \nu} R^{\mu \nu}&=& \frac{53747712 \lambda ^2 }{\left(32 \lambda +27 r^3\right)^6}\Big[2048 \lambda ^2\\
&+&3645 r^6-1728 \lambda  r^3\Big]\ .
 \end{eqnarray}
 
The square of Ricci tensor given by Eq.(\ref{RR1}) vanishes as the parameter $\gamma$ tends to zero and for the nonzero values of $\gamma$ parameter the square of Ricci tensor has the finite value as Ricci scalar: 
 \begin{eqnarray}\label{RR2}
  &&\lim_{r \to 0}R_{\mu \nu} R^{\mu \nu}=\frac{9216}{\gamma ^2 \lambda ^2 M^4\Omega_+^2}\ ,
 \\
&&\lim_{\gamma \to \infty} \left(\lim_{r \to 0}R_{\mu \nu} R^{\mu \nu}\right)=\frac{6561}{64 \lambda ^2 M^4}\ .
 \end{eqnarray}

{\bf The Kretschmann scalar. }
Explore now the Kretschmann scalar, which gives more information about the curvature of the spacetime (\ref{metric}), because the Kretschmann scalar is not vanishing even for Ricci flat spacetimes and helpful to study the properties of a given Ricci flat spacetime. The exact analytic expression for the Kretschmann scalar with the spacetime of the metric (\ref{metric}) is
\begin{eqnarray}\label{K}
\nonumber  K&=&R_{\mu \nu \sigma \rho}R^{\mu \nu \sigma \rho}=\frac{16 M^2}{\left(\lambda  M^2 \Omega_+  (\gamma  M+r)+r^3\right)^6}
\\\nonumber
 &&\times \Big[3 r^{12}-2 \lambda  M^2 r^9 \Omega_+  (6 \gamma  M+r)+\lambda ^2 M^4 r^6 \Omega ^2 
 \\\nonumber
 &&\times \left(54 \gamma ^2 M^2+48 \gamma  M r+13 r^2\right)+2 \lambda ^3 M^6 r^3 \Omega_+^3 
 \\\nonumber
 && \times \left(-3 \gamma ^3 M^3+6 \gamma ^2 M^2 r+7 \gamma  M r^2+2 r^3\right) \\
 && +\lambda ^4 M^8 \Omega_+^4 (6 \gamma ^4 M^4+16 \gamma ^3 M^3 r+19 \gamma ^2 M^2 r^2\nonumber \\ && +10 \gamma  M r^3+2 r^4)\Big], 
  \end{eqnarray}
which has the following limiting values:
 \begin{eqnarray}\label{K1}
 \nonumber
  \lim_{\gamma \to 0}K&=&\frac{16 M^2 }{r^2 \left(\lambda  M^2+r^2\right)^6} \Big[2 \lambda ^4 M^8+4 \lambda^3 M^6 r^2\\
 & +&13 \lambda ^2 M^4 r^4-2 \lambda  M^2 r^6+3 r^8\Big]\ ,
 \\\nonumber
  \lim_{\gamma \to \infty}K&=&\frac{34992 M^2}{\left(32 \lambda  M^3+27 r^3\right)^6} \Big[2097152 \lambda ^4 M^{12}
  \\\nonumber
 &-& 1769472 \lambda ^3 M^9 r^3
 +13436928 \lambda ^2 M^6 r^6
   \\
 &- & 2519424 \lambda  M^3 r^9
 +531441 r^{12}\Big]\ .
  \end{eqnarray}
The values of the Kretschmann scalar at the center of the BH spacetime take the following form:
\begin{eqnarray}\label{K2}
&&\lim_{r \to 0}K=\frac{6144}{\gamma ^2 \lambda ^2 M^4\Omega_+^2}, \\\nonumber
\\
&&\lim_{\gamma \to \infty}\left( \lim_{r \to 0}K\right)=\frac{2187}{32 \lambda ^2 M^4}. 
 \end{eqnarray} 
 
\begin{figure*}[ht!]
   \centering
  \includegraphics[width=0.323\linewidth]{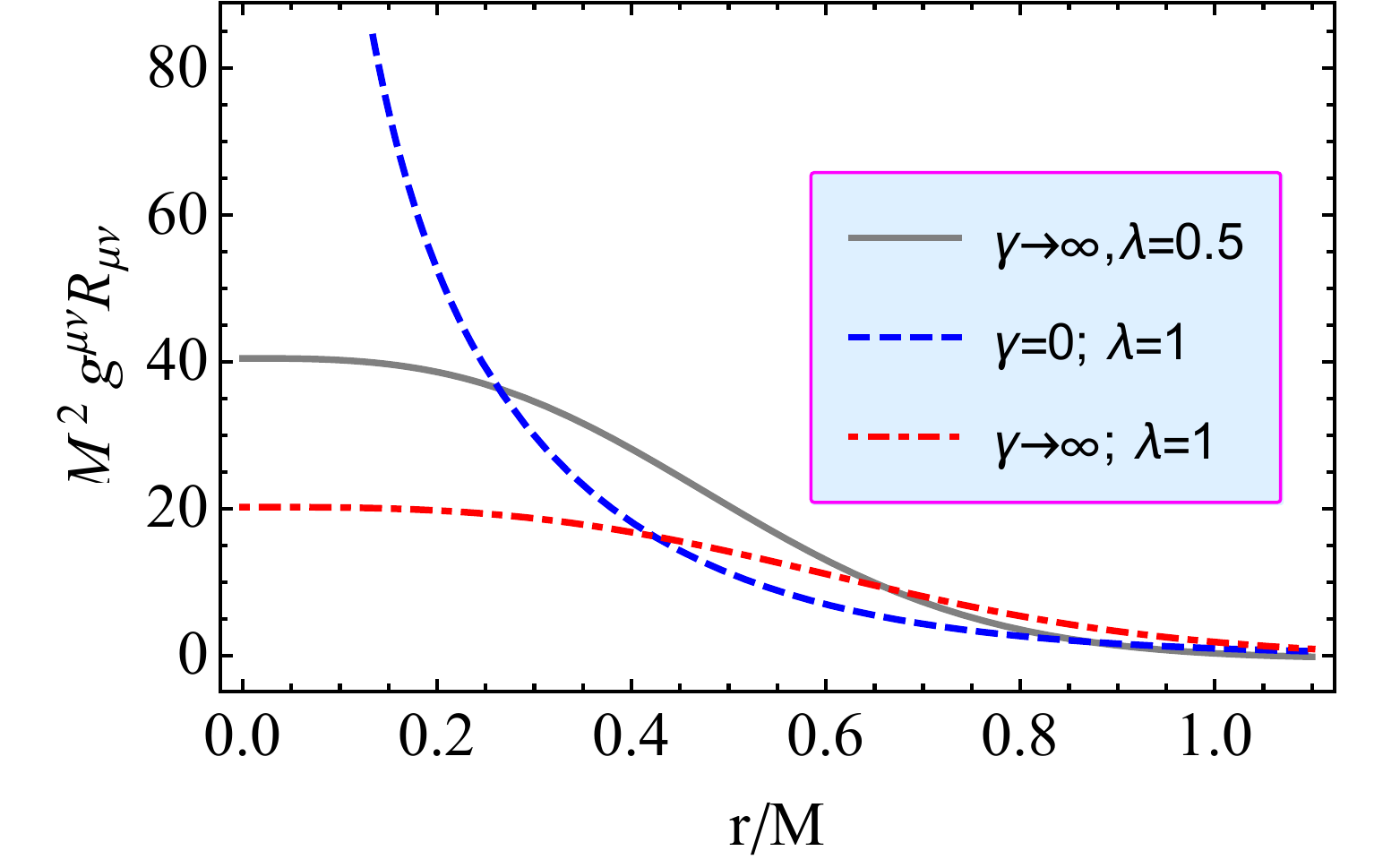}
   \includegraphics[width=0.323\linewidth]{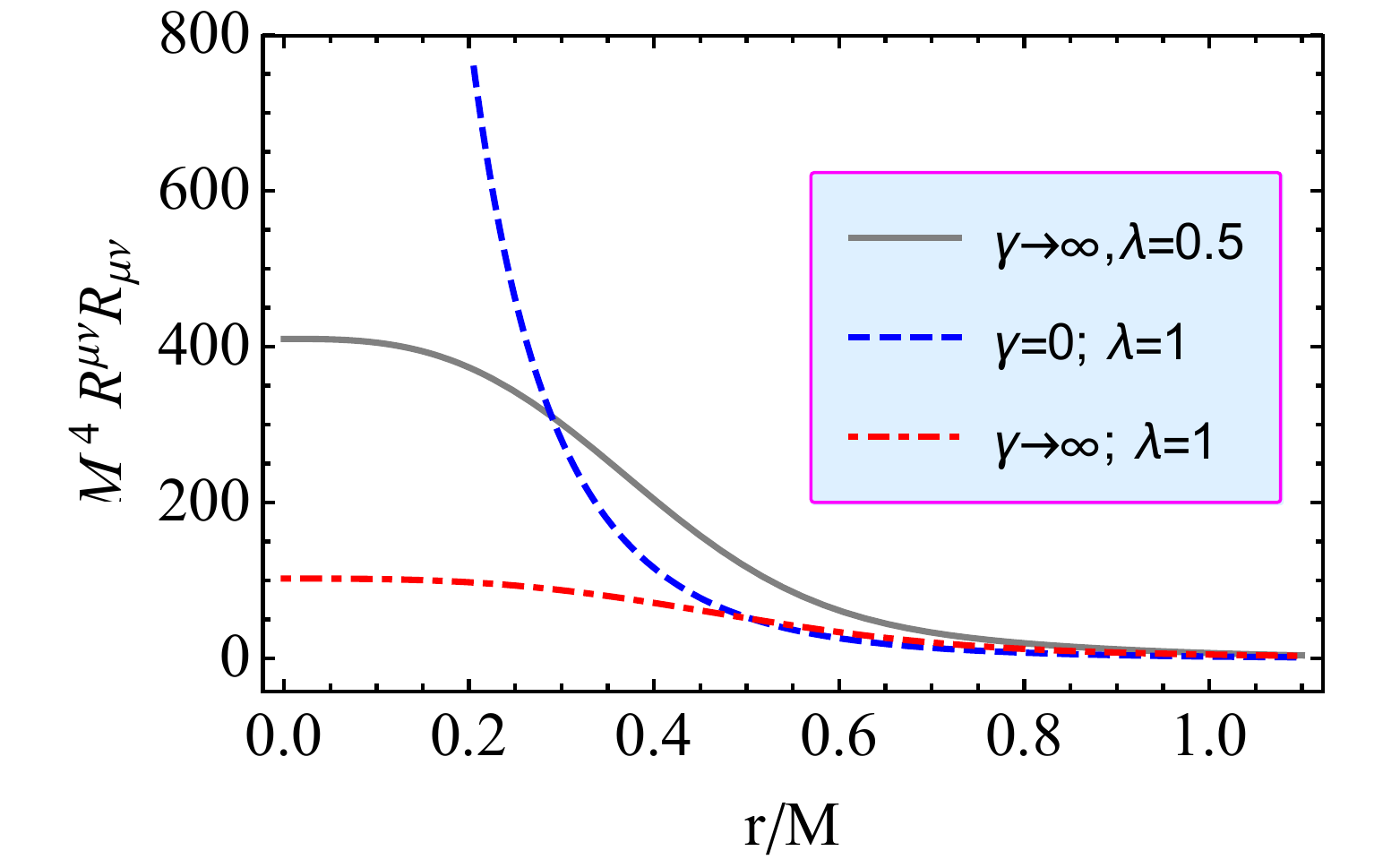}
    \includegraphics[width=0.323\linewidth]{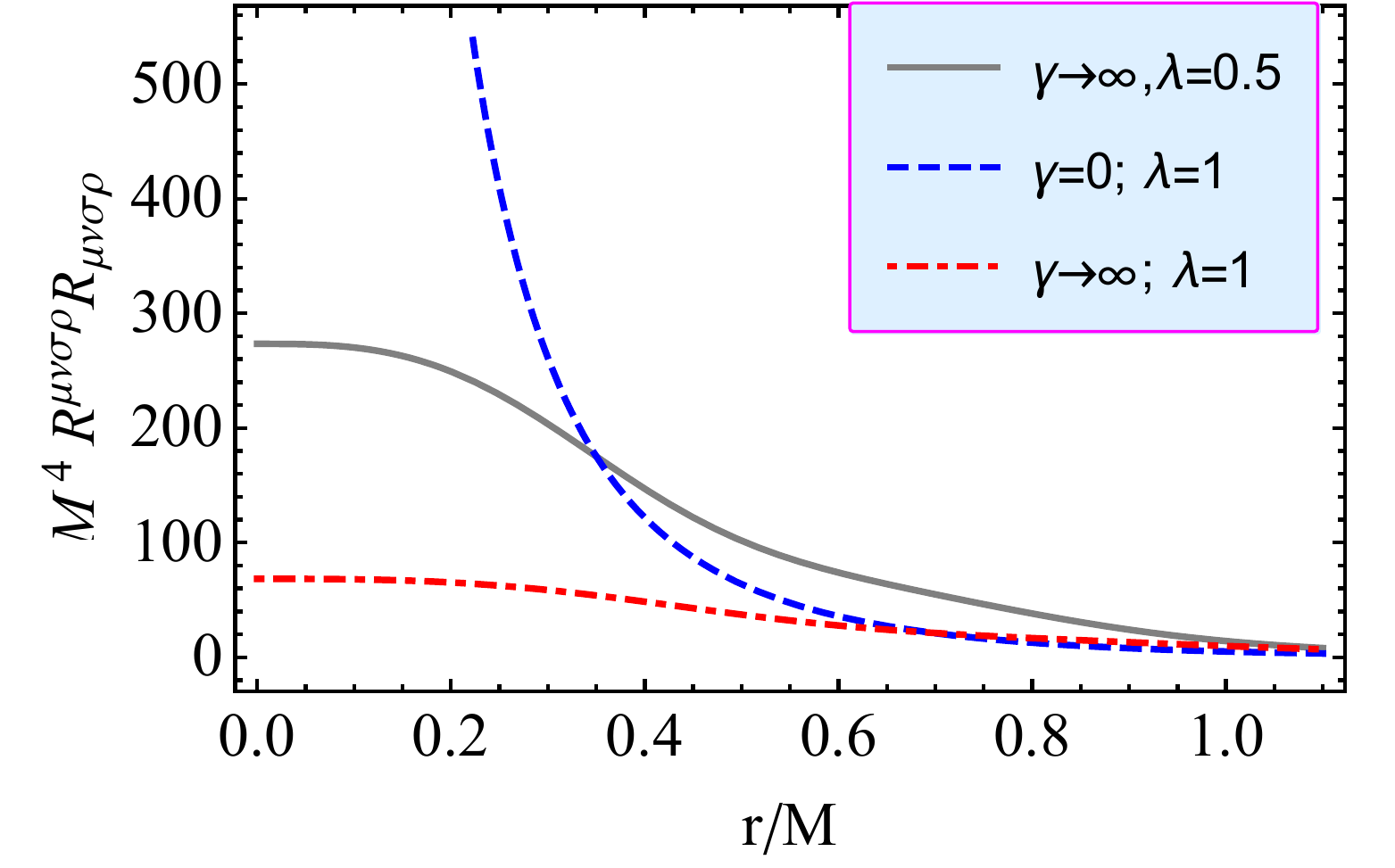}
	\caption{Radial profiles of scalar invariants of the spacetime (\ref{metric}) for the different values of parameters $\gamma$ and $\lambda$. Left panel corresponds to Ricci scalar, middle panel to square of Ricci tensor and right one to Kretchmann scalar. \label{ricci}}
\end{figure*} 

Figure~\ref{ricci} illustrates the radial dependence of scalar invariants for the different values of $\gamma$ and $\lambda$ parameters. One may see from Fig.~\ref{ricci} and Eqs.(\ref{R2}), (\ref{RR2}) and (\ref{K2}) that the increase of both parameters of RGI Schwarzschild BH $\gamma$ and $\lambda$ cause the decrease of the values of Ricci scalar, the square of Ricci tensor and the Kretschmann curvature scalar at the center of the RGI Schwarzschild BH. One may also see that when $\gamma=0$ the curvature scalars diverge at the center of the BH.

\subsection{Event horizon} 

Now we explore the event horizon properties of the RGI BH spacetime which governed by the lapse function (\ref{metfunct}) against to the parameters $\gamma$ and $\lambda$.

\begin{figure}[ht!]
   \centering
  \includegraphics[width=0.98\linewidth]{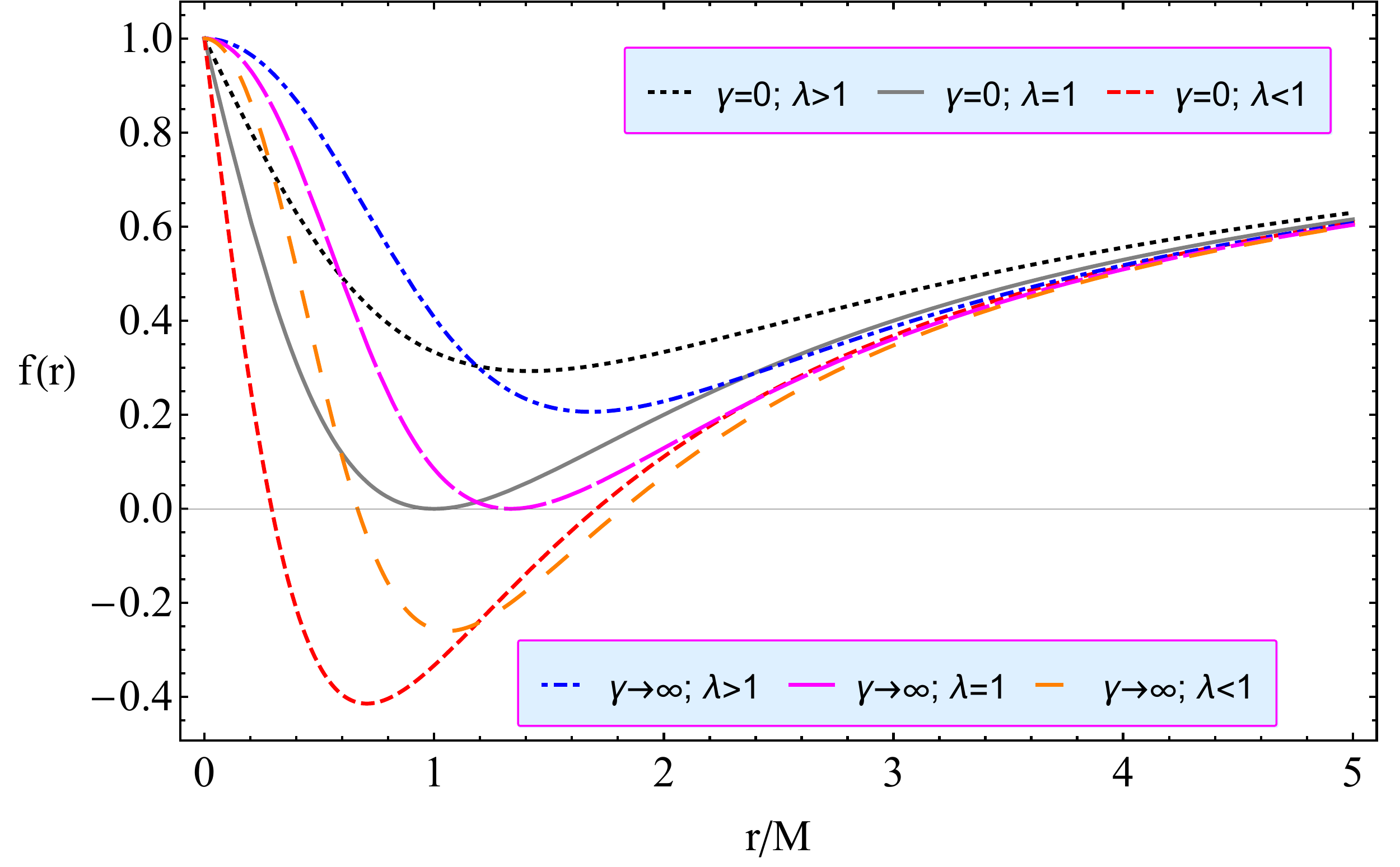}
	\caption{Dependence of lapse function from the radial coordinate $r/M$ for the different values of the parameters $\gamma$ and $\lambda$. \label{lapse}}
\end{figure} 

 Generally, the radius of the event horizon of a BH is described by the standard way setting $g_{rr} \to \infty, \ g^{rr}=0$ or equivalently, through the solution the equation 
 \begin{equation} \label{rhor}
     f(r)=0
 \end{equation}
 with respect to $r$ and calculations show that:
\begin{itemize}
    \item when $\lambda<1$ two event horizons do exist: inner and outer ones;
    \item in the case  when $\lambda>1$ the solution does not have any horizon;
    \item if $\lambda=1$ we have extreme RGI Schwarzschild BH and the two horizons coincide (see Fig.~\ref{lapse})  taking the following form
    \end{itemize}.

\begin{eqnarray}\label{evhoriz}
r_+\vert_{_{\lambda=1}}&=&r_-\vert_{_{\lambda=1}}=\frac{2}{3}\left(1+\frac{{\cal D}}{4\sqrt{2}}\right) +\frac{\sqrt[3]{2} (9 \gamma +2)}{12 {\cal D}}
\nonumber\\
&&\left(9 \gamma +10-3 \sqrt{\gamma +2} \sqrt{9 \gamma +2}\right)\ ,
\\
\nonumber
{\cal D}^3&=&(9 \gamma +2) \Big[81 \gamma ^2-9 \gamma  \left(3 \sqrt{\gamma +2} \sqrt{9 \gamma +2}-16\right)
\\
&-&18 \sqrt{\gamma +2} \sqrt{9 \gamma +2}+28\Big]\ . 
\end{eqnarray}

Using the real solutions of Eq.~\ref{rhor} one may see the behaviour of the event horizon against the variation of the parameters $\lambda$ and $\gamma$.

\begin{figure}[h!]
   \centering
  \includegraphics[width=0.98\linewidth]{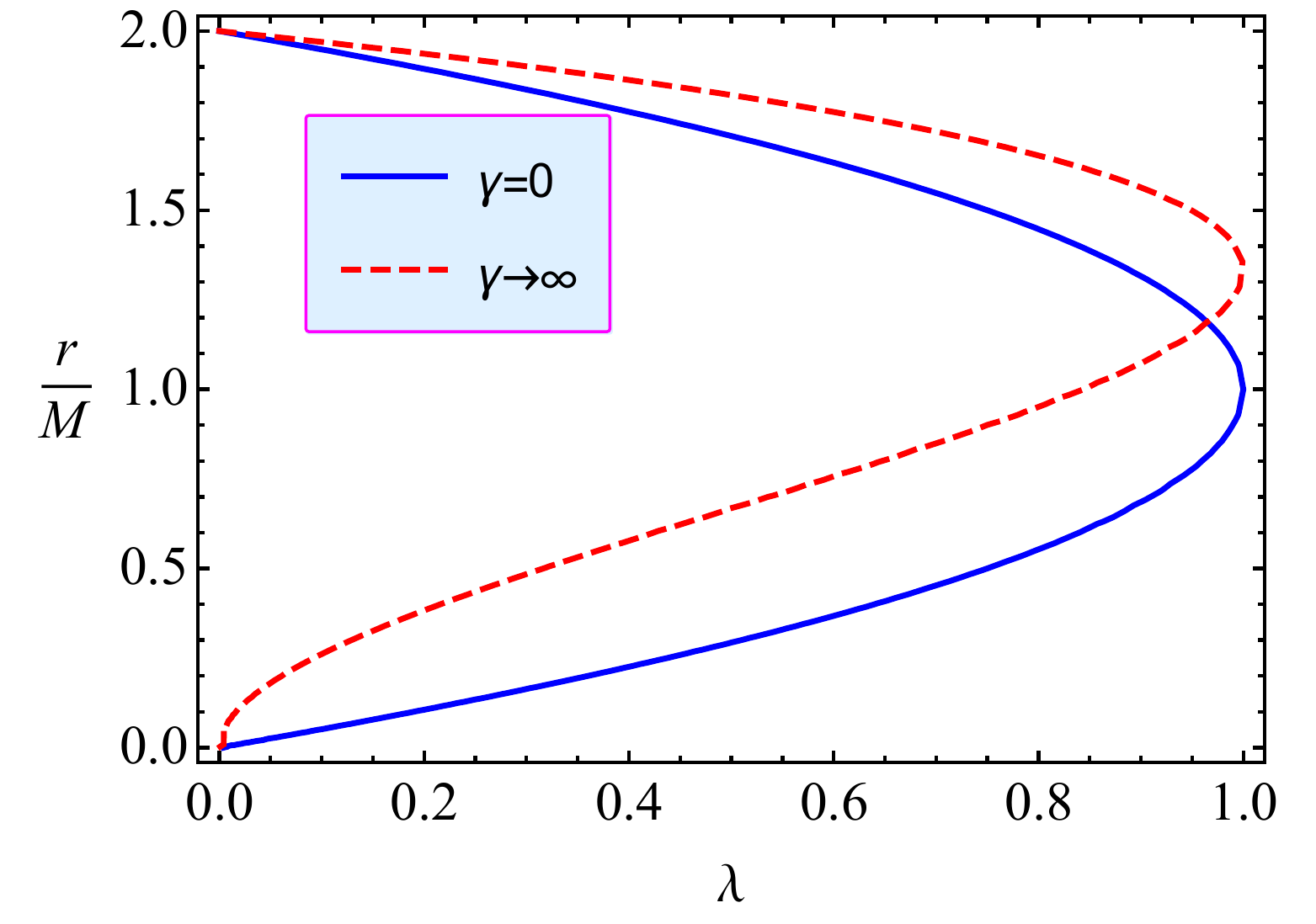}
	\caption{Dependence of event horizon radius on parameter $\lambda$ for the different values of the parameter $\gamma$. \label{horizon}}
\end{figure} 

The dependence of radius of event horizon from parameter $\lambda$ for the different values of $\gamma$ has been shown in Fig.~\ref{horizon}. One can see that with the increase of $\lambda$ parameter outer horizon decreases and inner horizon increases and merge each other at $\lambda=1$ and there is no horizon out of $\lambda=1$. Moreover, the increase of the parameter $\gamma$ for the given value of the parameter $\lambda$ increases the outer and inner event horizons.

In the limiting cases when $\gamma\to 0$ and $\gamma \to \infty $ the event horizon described by~(\ref{evhoriz}) takes the following values 
\begin{eqnarray}
\lim_{\gamma\to 0}r_-\vert_{ _{\lambda=1}}&=&\lim_{\gamma\to 0}r_+\vert_{ _{\lambda=1}}=M\ ,
\\
\lim_{\gamma\to \infty}r_-\vert_{ _{\lambda=1}}&=&\lim_{\gamma\to \infty}r_+\vert_{ _{\lambda=1}}=\frac{4}{3}M\ ,
\end{eqnarray}
which indicate that event horizon at $\lambda=1$ can take place at the values in the range of $r_h/M \in (1,4/3)$ corresponding to the values of the parameter $\gamma \in (0; \infty)$.

\section{Test particle motion \label{testpartmotion}}
 
Here we investigate in detail the motion of test particles in the spacetime~(\ref{metric}).

\subsection{Equation of motion} 
Consider the dimensionless Lagrangian density for a neutral particle with mass $m$
\begin{eqnarray}
\mathscr{L}_{\rm p}=\frac{1}{2} g_{\mu\nu} \dot{x}^{\mu} \dot{x}^{\nu} ,
\end{eqnarray}
then the conserved quantities of motion read
\begin{eqnarray}
\label{consts1}
&& p_t=\frac{\partial \mathscr{L}_{\rm p}}{\partial \dot{t}}\Longrightarrow g_{tt}\dot{t}=-{\cal E}\ ,
\\\label{consts2}
&& p_{\phi}=\frac{\partial \mathscr{L}_p}{\partial \dot{\phi}} \Longrightarrow g_{\phi \phi}\dot{\phi} = l\ ,
\end{eqnarray}
where ${\cal E}=E/m$ and  $l=L/m$  are specific energy and angular momentum of the particle, respectively. Equations of motion for a test particle  then governed by the normalization condition
\begin{equation}\label{norm4vel}
g_{\mu \nu}u^{\mu}u^{\nu}=\epsilon \ ,
\end{equation}
where $\epsilon$ has the values $0$ and $-1$ for massless and massive particles, respectively.

For the massive neutral particles the motion is governed by timelike geodesics of the spacetime (\ref{metric}) and the equations of motion can be found by using Eq.(\ref{norm4vel}). Taking into consideration equations (\ref{consts1})-(\ref{consts2}), we obtain the equations of motion in the separated and integrated form
\begin{eqnarray}\label{eqmotionneutral}
\dot{r}^2&=&{\cal E}^2+g_{tt}\left(1+\frac{\cal K}{r^2}\right)\ ,
 \\
\dot{\theta}&=&\frac{1}{g_{\theta \theta}^2}\Big({\cal K}-\frac{l^2}{\sin^2\theta}\Big)\ ,
 \\
\dot{\phi}&=&\frac{l}{g_{\phi \phi}}\ ,
 \\
\dot{t}&=&-\frac{{\cal E}}{g_{tt}}\ ,
\end{eqnarray}
where ${\cal K}$ denotes the Carter constant corresponding to the total angular momentum.

Restricting the motion of the particle to the constant plane, in which $\theta=\rm const$ and $\dot{\theta}=0$, the Carter constant takes the form ${\cal K}=l^2/\sin^2\theta$ and the equation of the radial motion can be expressed in the form

\begin{eqnarray}
 \dot{r}^2={\cal E}^2-V_{\rm eff}\ ,
\end{eqnarray}
where the effective potential of the motion of neutral particles reads
\begin{eqnarray}\label{effpotentail}
V_{\rm eff} = f(r)\left(1+\frac{l^2}{r^2\sin^2\theta}\right) .
\end{eqnarray}

Now one may consider the conditions for the circular motion, that means no radial motion ($\dot{r}=0$) and no forces are in the radial direction $\ddot{r}=0$) and obtain  the radial profiles of the specific angular momentum and specific energy for circular orbits at the equatorial plane ($\theta=\pi/2$) in the following form
\begin{eqnarray}\label{LandE}
l= \frac{r^3 \mathcal{Z}}{2-r \mathcal{Z}} \ , \qquad {\cal E}=\frac{2 f(r)}{2-r \mathcal{Z}}\ ,
\end{eqnarray}
where  ${\cal Z} = \partial_r\ln f(r)$. 

\begin{figure}[h!]
   \centering
  \includegraphics[width=0.98\linewidth]{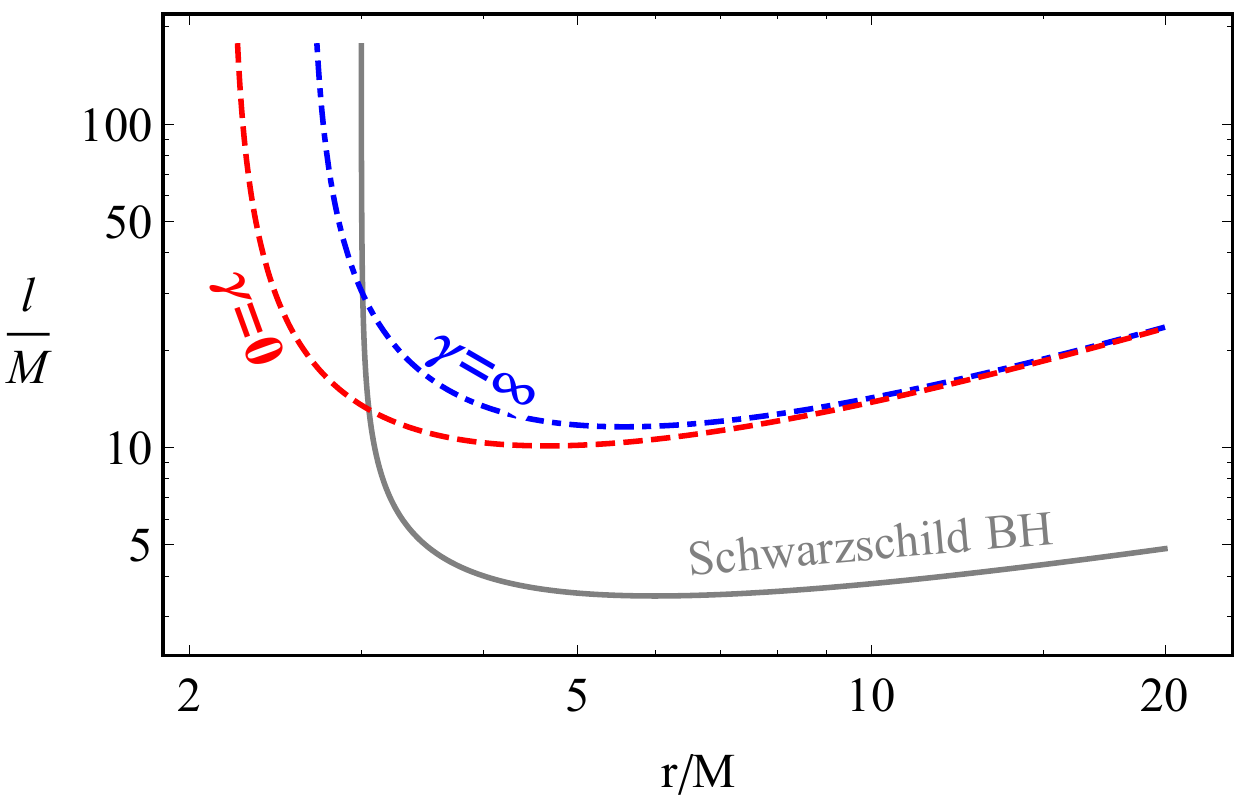}
  \includegraphics[width=0.98\linewidth]{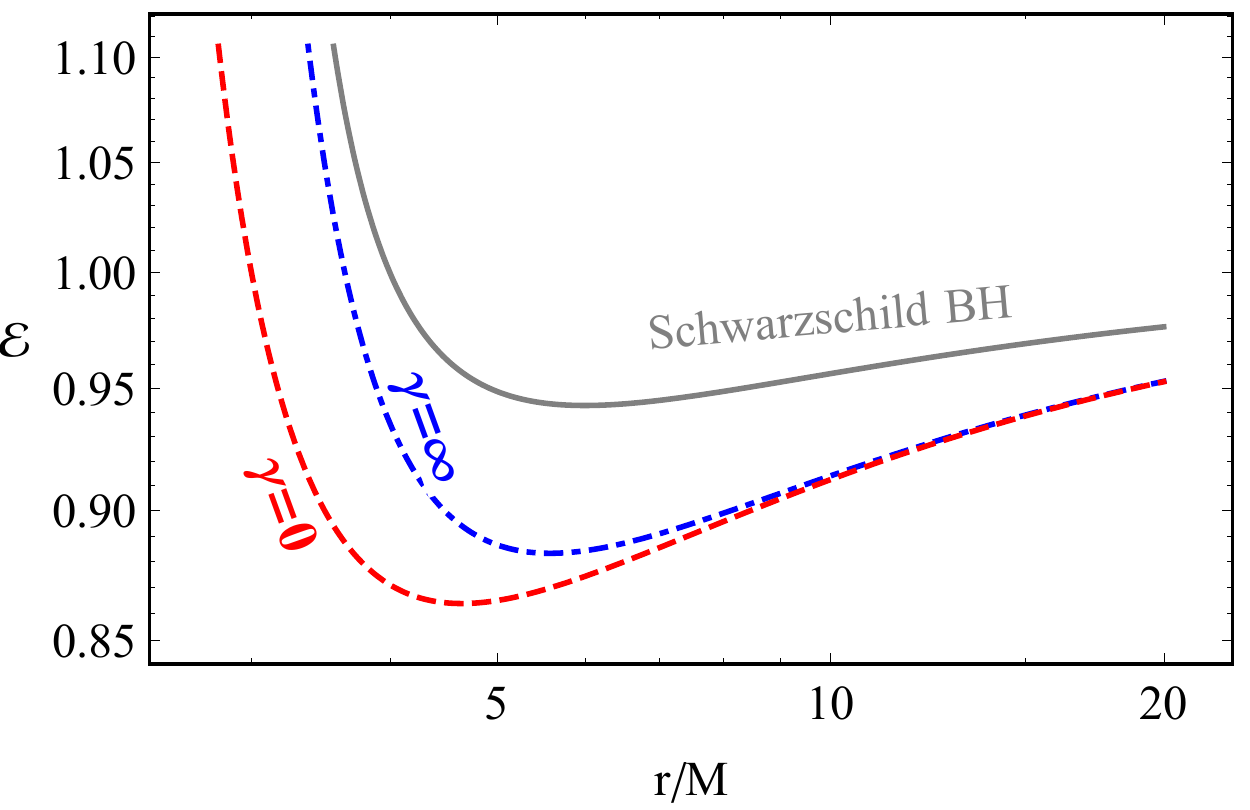}
	\caption{Dependence of specific angular momentum (on the top panel) and energy (at the bottom panel) for circular orbits from the radial coordinate for the different values of $\gamma$ at the fixed value of the parameter $\lambda=1$ with comparison to the Schwarzschild case. \label{landefign}}
\end{figure}

Figure~\ref{landefign} demonstrates the radial dependence of specific energy and angular momentum for the values of the parameter $\gamma=0$ and $\gamma \to \infty$ at the fixed values of the parameter $\gamma=1$ with respect to the Schwarzschild BH. From the top panel of Fig.~\ref{landefign} one may conclude that the specific angular momentum increases with the increase of both parameters $\lambda$ and $\gamma$, while the specific energy decreases.

\subsection{Innermost stable circular orbits}

The innermost stable circular orbits (ISCO) can be defined by the solution of equation governed by the condition $\partial_{rr}V \geq 0$:
\begin{eqnarray}\label{iscoeq}
\mathcal{Z} f(r) (2 r \mathcal{Z}-3)\geq r f''(r)\ ,
\end{eqnarray}
where the prime $'$ denotes partial derivative with respect to the radial coordinate. 

It is impossible to find the exact solutions of Eq.(\ref{iscoeq}), however, the limits of the solution at the fixed value of the parameter $\lambda=1$ can be obtained in the following form 
\begin{equation}\label{iscolimits}
\lim_{\gamma \to \infty}r_{\rm isco}=5.58396\rm M, \qquad \lim_{\gamma \to 0}r_{\rm isco}=4.64575\rm M \ .
\end{equation}
It implies that the increase of the parameter $\lambda$ ($\gamma$) causes to decrease (increase) ISCO radius and minimum value of ISCO radius is $4.64575\rm M $ and the maximum one is $5.58396\rm M$. 

\subsection{RGI Schwarzschild BH vs Kerr BH}

In this subsection we will explore the values of the parameter $\lambda$ which can mimic the spin of Kerr BH providing the same ISCO radius. The problem is that both rotating Kerr BH and RGI Schwarzchild BH parameters cause to increase ISCO radius. Here, we plan to show how to distinguish the BHs through (direct or indirect) measurements of ISCO radius. 
ISCO radius of the test particles for retrograde and prograde orbits  around Kerr BH can be expressed as ~\cite{Bardeen72}
\begin{eqnarray}
r_{\rm isco}= 3 + Z_2 \pm \sqrt{(3- Z_1)(3+ Z_1 +2 Z_2 )} \ ,
\end{eqnarray}
with
\begin{eqnarray} \nonumber
Z_1 &  = & 
1+\left( \sqrt[3]{1+a}+ \sqrt[3]{1-a} \right) 
\sqrt[3]{1-a^2} \ ,
\\ \nonumber
Z_2 & = & \sqrt{3 a^2 + Z_1^2} \ .
\end{eqnarray}

\begin{figure}[ht!]
   \centering
  \includegraphics[width=0.98\linewidth]{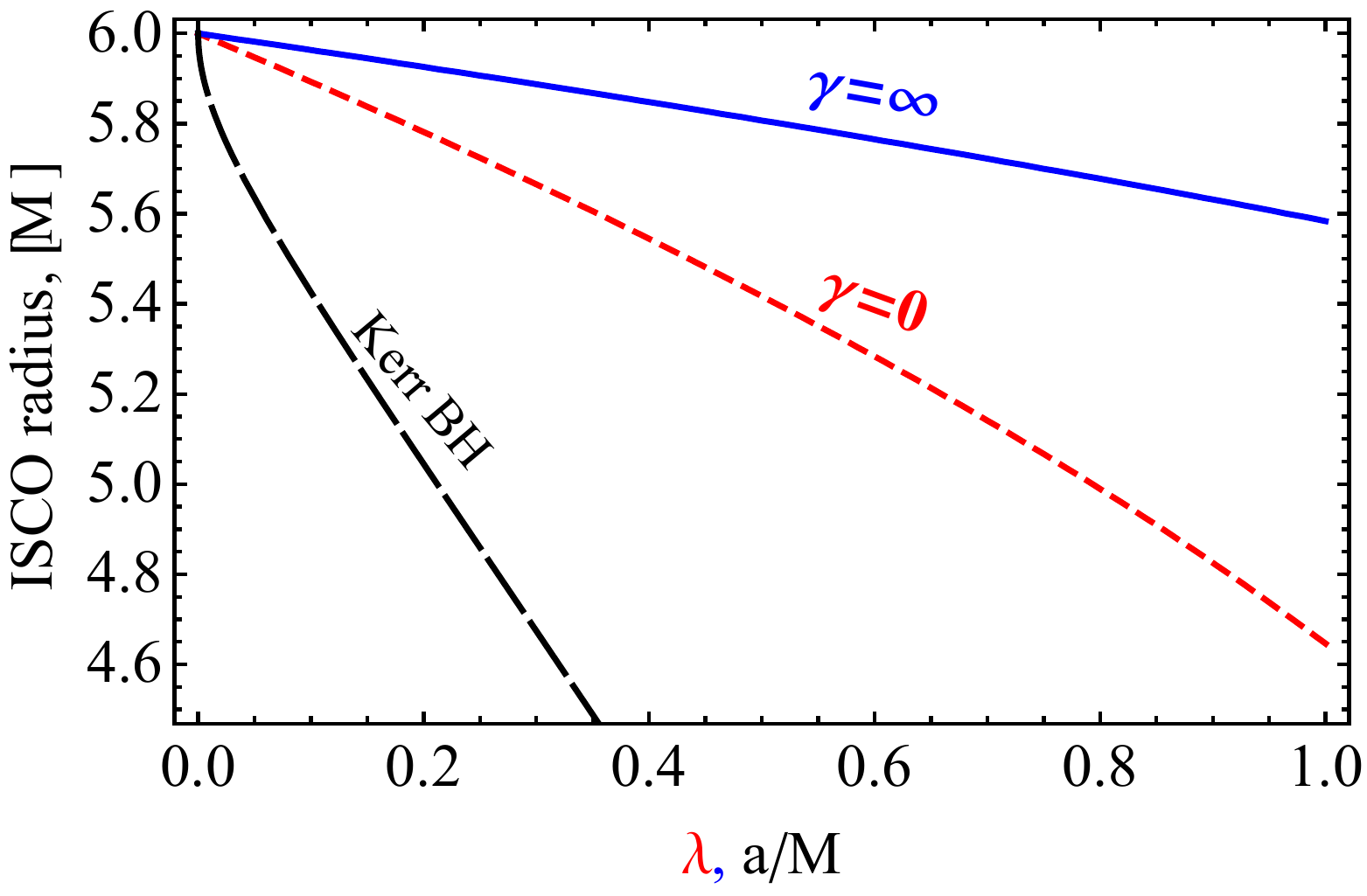}
	\caption{Dependence of ISCO radius from parameter $\lambda$ for the different values of the parameter $\gamma$. \label{isco}}
\end{figure} 

Figure~\ref{isco} shows ISCO profiles of Kerr BH and RGI Schwarzchild BH at the range of the parameter $\gamma$ from zero up to infinity. One may see from the figure that the effect of Kerr BH spin much stronger than that of  the parameters of the RGI Schwarzschild BH. 

Now we are back to main goal of this section which is to determine values of spin parameter and the parameters $\gamma$ and $\lambda$ which may provide the same ISCO radius.
\begin{figure}[h!]
   \centering
  \includegraphics[width=0.9\linewidth]{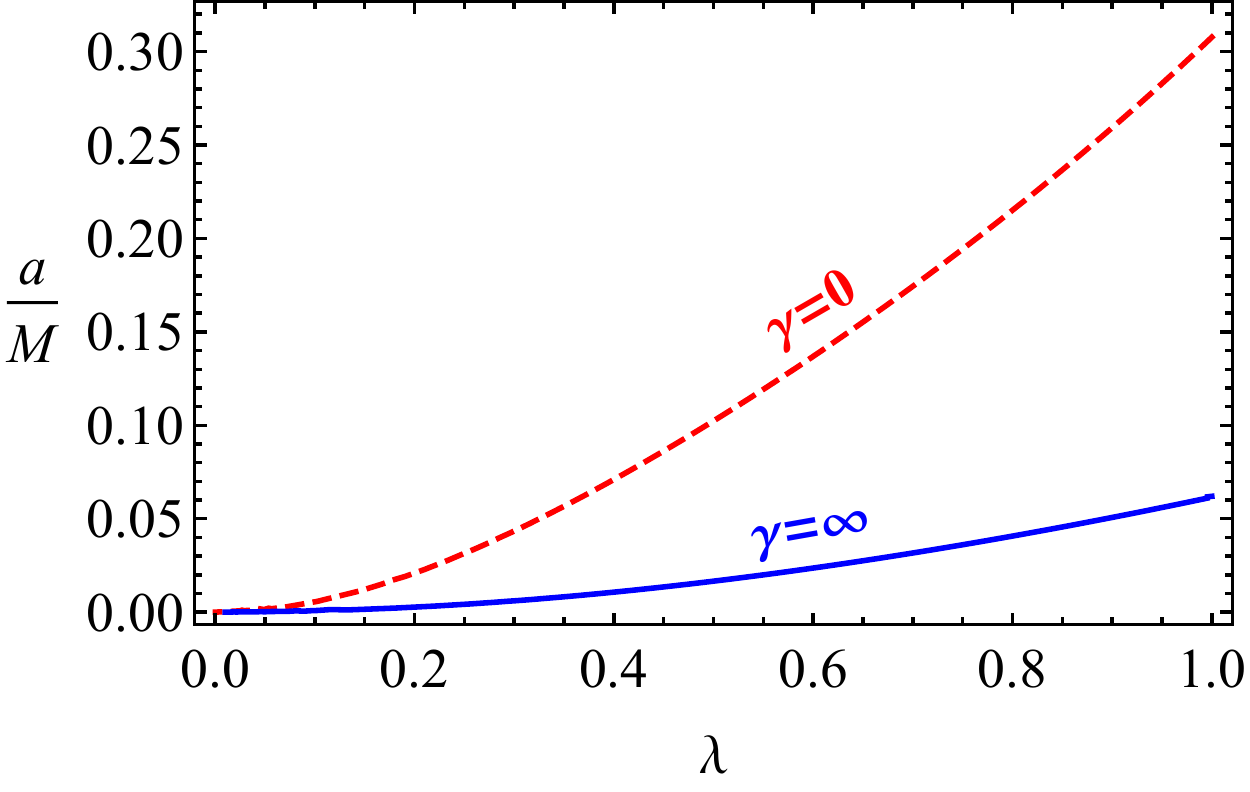}
	\caption{The degeneracy graphs between spin and $\lambda$ parameters providing the same ISCO radius for the different values of the parameter $\gamma$ \label{mimic1}}
\end{figure} 

Figure~\ref{mimic1} illustrates relations between spin and the $\lambda$ parameters giving the same ISCO radius for test particles. One can see from the figure that the patameter $\lambda$ can mimic BH spin parameter up to $a/M=0.307681$  in the case when $\gamma=0$, while at $\gamma \to \infty$ it can mimic up to $a/M = 0.0615279$.
Since, according to the accretion disks observations the astrophysical black holes are rapidly rotating wich the spin parameter up to $a/M \simeq 0.99$ the neutral particle motion can not mimic the black hole spin.

\subsection{The energy efficiency}

\begin{figure}[ht!]
   \centering
  \includegraphics[width=0.98\linewidth]{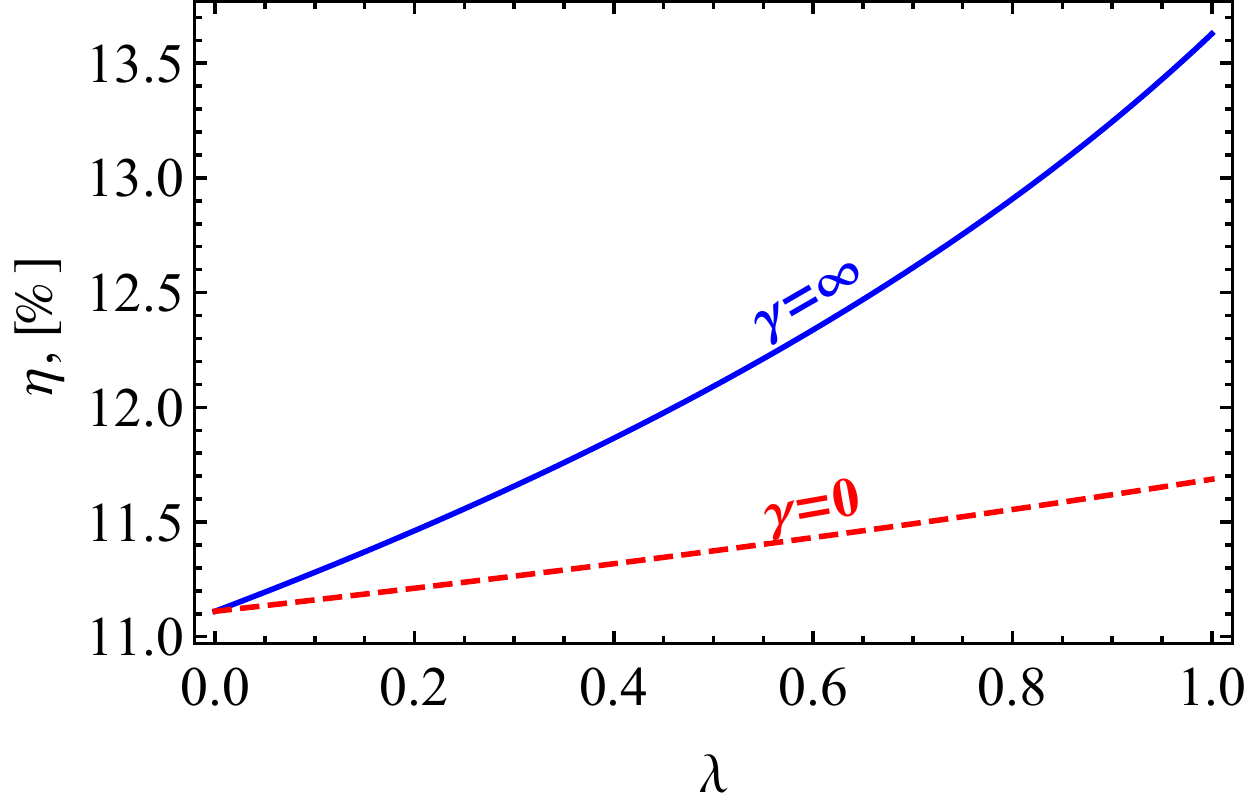}
	\caption{Dependence of energy efficiency from parameter $\lambda$ for the different values of the parameter $\gamma$ \label{efficiency}}
\end{figure}

According to Novikov-Thorn model the Keplerian accretion around an astrophysical BH is modeled as the geometrically thin disks which governed by the properties of the spacetime circular geodesics~\cite{Novikov73}. Generally, the energy efficiency of the accretion disk around a BH means the maximum energy can be extracted as radiation of the falling matter in to the central BH from the disk.  The efficiency of the accretion of the test particle can be calculated through the following expression
\begin{equation}
\eta=1-{\cal E}\,  \vline_{\,r=r_{\rm ISCO}},
\end{equation}
where ${\cal E}_{\rm ISCO}$ is the energy of the particle at the ISCO which is characterized by the ratio of the binding energy (BH- particle system) and rest energy of test particle and it can be calculated using the energy of the particles given in Eq.(\ref{LandE})  at ISCO. The exact  form of the efficiency is hard to calculate analytically due to complicated form of ISCO radius and we show the effects of RGI Schwarzschild BH parameters on the efficiency in graphical form in Fig.~\ref{efficiency}.

One can see that the value of the energy efficiency is increased with the increase of both parameters of the RGI chwarzschild BH and as the effect of the increase of the parameter $\gamma$ from zero to infinity the efficiency increases to 2 \% and the increase of the parameter $\lambda$ from zero up to 1 increases to about 3.5 \%. It implies that the quantum effects on accretion disc radiations are not reasonable and not measurable.

\section{Dynamics of charged particles \label{chargedpartmotion}}

\subsection{Electromagnetic four-potentials}

Since, the spacetime of RGI Schwarschild black hole Ricci flat outside of outer horizon (see subsection \ref{RRKinvariants}), the electromagnetic four-potential around the RGI Schwarzschild BHs immersed in an external asymptotically uniform magnetic field with the asymptotic value $B_0$, can be found using conserved timelike and spacelike Killing vectors with the help of the Wald method ~\cite{Wald74} being valid for Ricci flat spacetime in the following form , assuming the magnetic field is perpendicular to the equatorial plane where $\theta =\pi/2$.
\begin{eqnarray}\label{Aft}
A_{\phi} & = & \frac{1}{2}B_0 r^2\sin^2\theta,
\\\nonumber
A_t & = & 0 =A_r=A_{\theta}\ ,
\end{eqnarray}
One may immediately obtain the non zero components of the electromagnetic field tensor by the relation ${\cal F}_{\mu\nu}=A_{\nu,\mu}-A_{\mu,\nu}$ in the standard form
\begin{eqnarray}\label{FFFF}
{\cal F}_{r \phi}&=&B_0 r\sin^2\theta \ ,
 \\
 {\cal F}_{\theta \phi}&=&B_0r^2\sin\theta \cos\theta\ .
\end{eqnarray}

We will calculate the orthonormal components of the magnetic field around the BH using the relation
\begin{eqnarray}\label{fields}
B^{\alpha} &=& \frac{1}{2} \eta^{\alpha \beta \sigma \mu} F_{\beta \sigma} u_{\mu}\ ,
\end{eqnarray}
$\eta_{\alpha \beta \sigma \gamma}$ is the pseudo-tensorial form of the Levi-Civita symbol $\epsilon_{\alpha \beta \sigma \gamma}$ with the relations 
\begin{eqnarray}
\eta_{\alpha \beta \sigma \gamma}=\sqrt{-g}\epsilon_{\alpha \beta \sigma \gamma}\, \qquad \eta^{\alpha \beta \sigma \gamma}=-\frac{1}{\sqrt{-g}}\epsilon^{\alpha \beta \sigma \gamma}\ ,
\end{eqnarray}
with $g={\rm det|g_{\mu \nu}|}=-r^4\sin^2\theta$ for spacetime metric (\ref{metric}) and 
\begin{eqnarray}
\epsilon_{\alpha \beta \sigma \gamma}=\begin{cases}
+1\ , \rm for\  even \ permutations
\\
-1\ , \rm for\  odd\  permutations
\\
\, \, 0\ , \, \rm for\ the\ other\ combinations
\end{cases}\ ,
\end{eqnarray}
and we have

\begin{equation}\label{BrBt}
    B^{\hat{r}}=B_0 \cos\theta, \\ \qquad B^{\hat{\theta}}=\sqrt{f(r)}B_0\sin \theta\ .
    \end{equation}
 Eq.(\ref{BrBt}) implies that the azimthal component of the magnetic field around BH is modified under the effect of gravitational field. However, the radial component of the magnetic field is formally as the Newtonian one.
    
\begin{figure}[h!]
    \centering
  \includegraphics[width=0.97\linewidth]{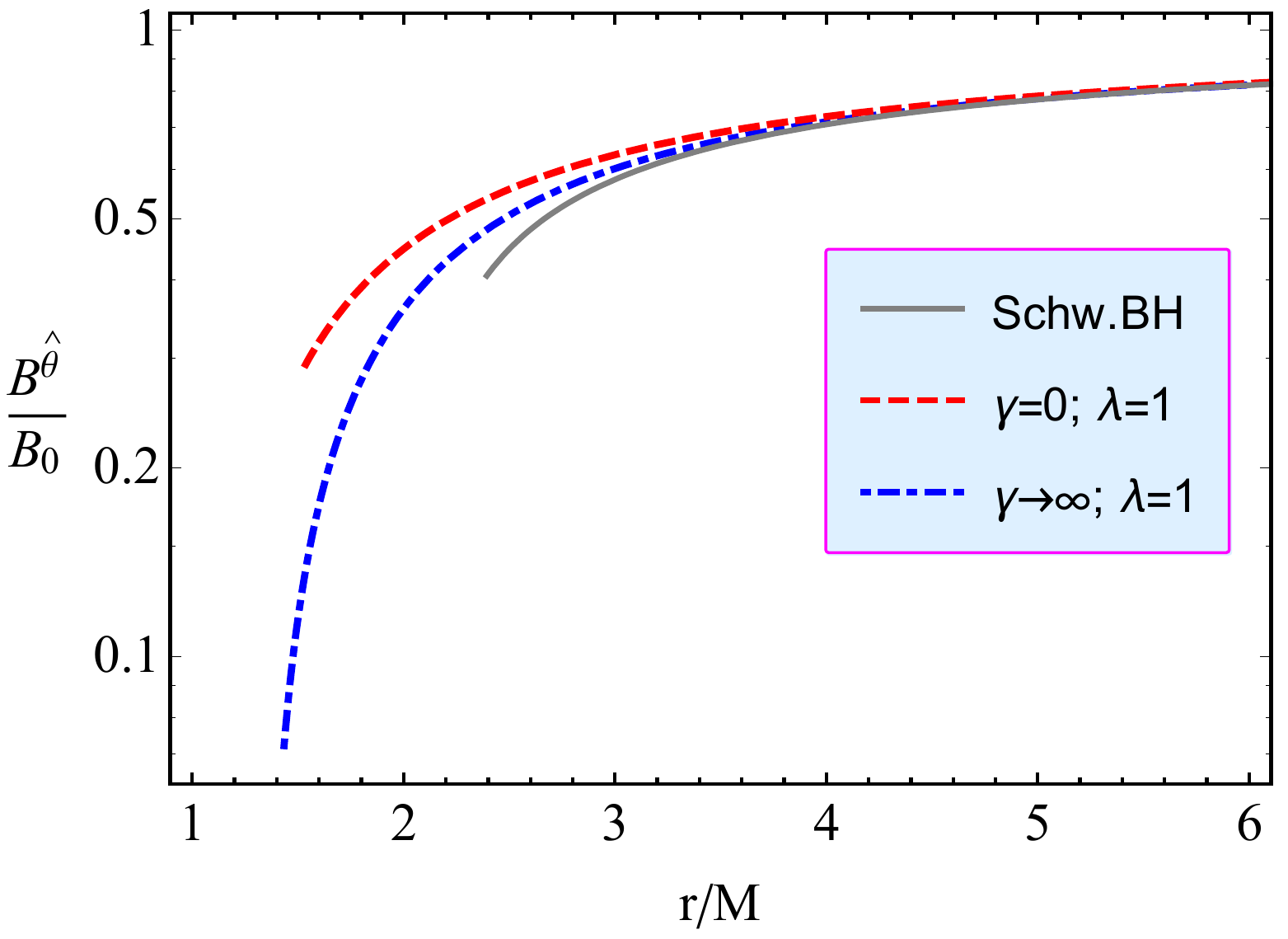}
    \caption{The radial dependence of azimuthal component of the external magnetic field $B^{\hat{\theta}}$  normalized to its asymptotic value $B_0$.}
    \label{Bt}
\end{figure}

The radial dependence of the azimuthal component of the external magnetic field around RGI Schwarzschild BHs is shown in Fig.~\ref{Bt} for the different values of the parameter $\gamma$. One can see that the magnetic field strongly depends on the parameter $\gamma$ and the increase of $\lambda$ parameter causes the increase of the magnetic field while the increase of $\gamma$ parameter forces to essentially decrease it.

\subsection{Equation of motion}
 Here we will study the motion of a charged particle with mass $m$ and electric charge $q$ around RGI Schwarzschild BH using the Hamilton-Jacobi equation
\begin{eqnarray}\label{HJ}
g^{\mu \nu}\Big(\frac{\partial {\cal S}}{\partial x^{\mu}}-q A_{\mu}\Big)\Big(\frac{\partial {\cal S}}{\partial x^{\nu}}-q A_{\nu}\Big)=-m^2\ .
\end{eqnarray}
Since, $t$ and $\phi$ are the Killing variables, the action can be written in the following form
\begin{eqnarray}\label{action}
{\cal S}=-{\cal{E}} t+L \phi+{\cal S}_r(r)+{\cal S}_{\theta}(\theta) \ .
\end{eqnarray}

One can find the expressions for the components of four velocity of the charged particle at equatorial plane in the following separable form:
\begin{eqnarray}\label{eqmotionch}
\nonumber
\dot{t}&=&\frac{{\cal E}}{f(r)}\ ,
\\
\nonumber
\dot{r}^2&=&{\cal E}^2-f(r)\Big[1+\Big(\frac{l}{r}-\omega_{\rm B} r\Big)^2\Big]\ ,
\\
\dot{\phi}&=&\frac{l}{r^2}-\omega_{\rm B}\ ,
\end{eqnarray}
where $\omega_{\rm B}=qB_0/(2mc)$ is the cyclotron frequency which corresponds to interaction of magnetic field and charged particle

\begin{eqnarray}
\dot{r}^2={\cal E}^2-V_{\rm eff} \ .
\end{eqnarray}

One can easily find effective potential for the particle by putting Eq.(\ref{action}) into Eq.(\ref{HJ}) in the following simple form
\begin{equation}
    V_{\rm eff}= f(r)\left[1+\left(\frac{l}{r \sin \theta}-\omega_B r \sin\theta\right)^2\right] \ .
\end{equation}

%%%%%%%%%%%%%%%%%%%%%%%%%%%%%%%%%%%%%
 \subsection{ISCO}
 
Stable circular orbits in the equatorial plane can be studied using the following standard conditions

\begin{eqnarray}\label{iscocond}
V_{\text{eff}}={\cal E}\ ,
\qquad
V'_{\text{eff}}=0\ ,
\qquad
V''_{\text{eff}} \geq 0\ .
\end{eqnarray}
The solutions of the equation $V'_{\text{eff}}=0$ in the fixed background imply that the particle with given specific charge $q$ follows a circular orbit at a given radius $r$, if its specific angular momentum is given by the relation

 \begin{eqnarray}\label{lpmeq}
 l_{\pm}= \frac{r}{{\cal Z}r-2}\left( {\cal Q}\pm \omega_{\rm B}{\cal Z} \right)\ ,
 \end{eqnarray}
 with ${\cal Q}^2=2{\cal Z}r+r^2 \left(4\omega_{\rm B}^2-{\cal Z}^2\right)$ and

 \begin{eqnarray}\label{epmeq}
 {\cal E}_{\pm} &=& \frac{2f(r)}{{\cal Z}r-2}\Big(2-{\cal Z}r+4\omega^2_{\rm B}r^2 \pm  \omega_{\rm B} r {\cal Q} \Big)\ .
 \end{eqnarray}
 
 Equations (\ref{lpmeq}) and (\ref{epmeq}) show that the values of the specific angular momentum and energy of the charged particles $l$ and ${\cal E}$ have symmetry by replacement of $\omega_{\rm B} \to -\omega_{\rm B}$ that mean $(l,{\cal E})_{-}(\omega_{\rm B}>0)=(l,{\cal E})_{+}(\omega_{\rm B}<0)$.
 
 \begin{figure}[ht!]
   \centering
  \includegraphics[width=0.98\linewidth]{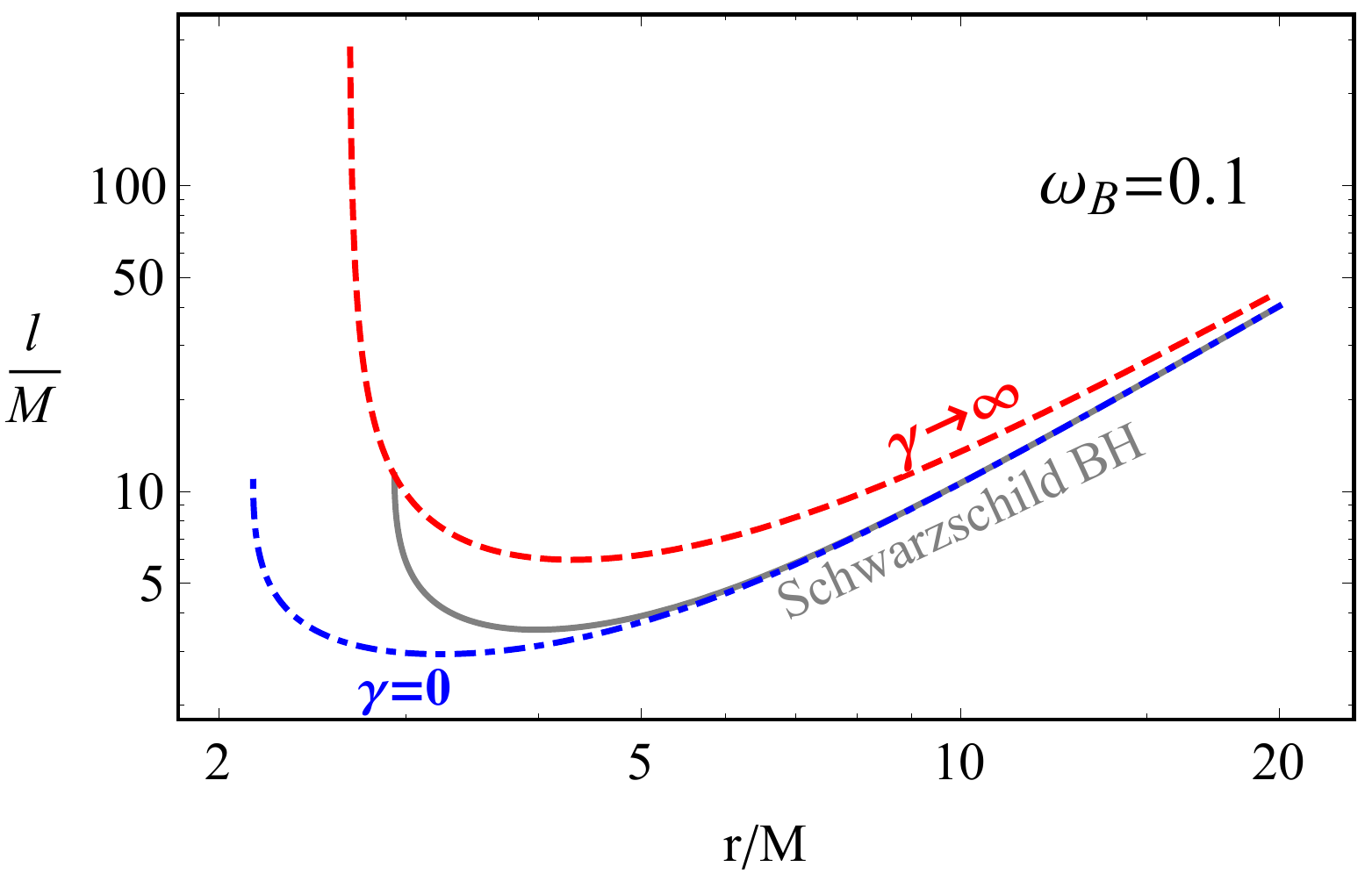}
  \includegraphics[width=0.98\linewidth]{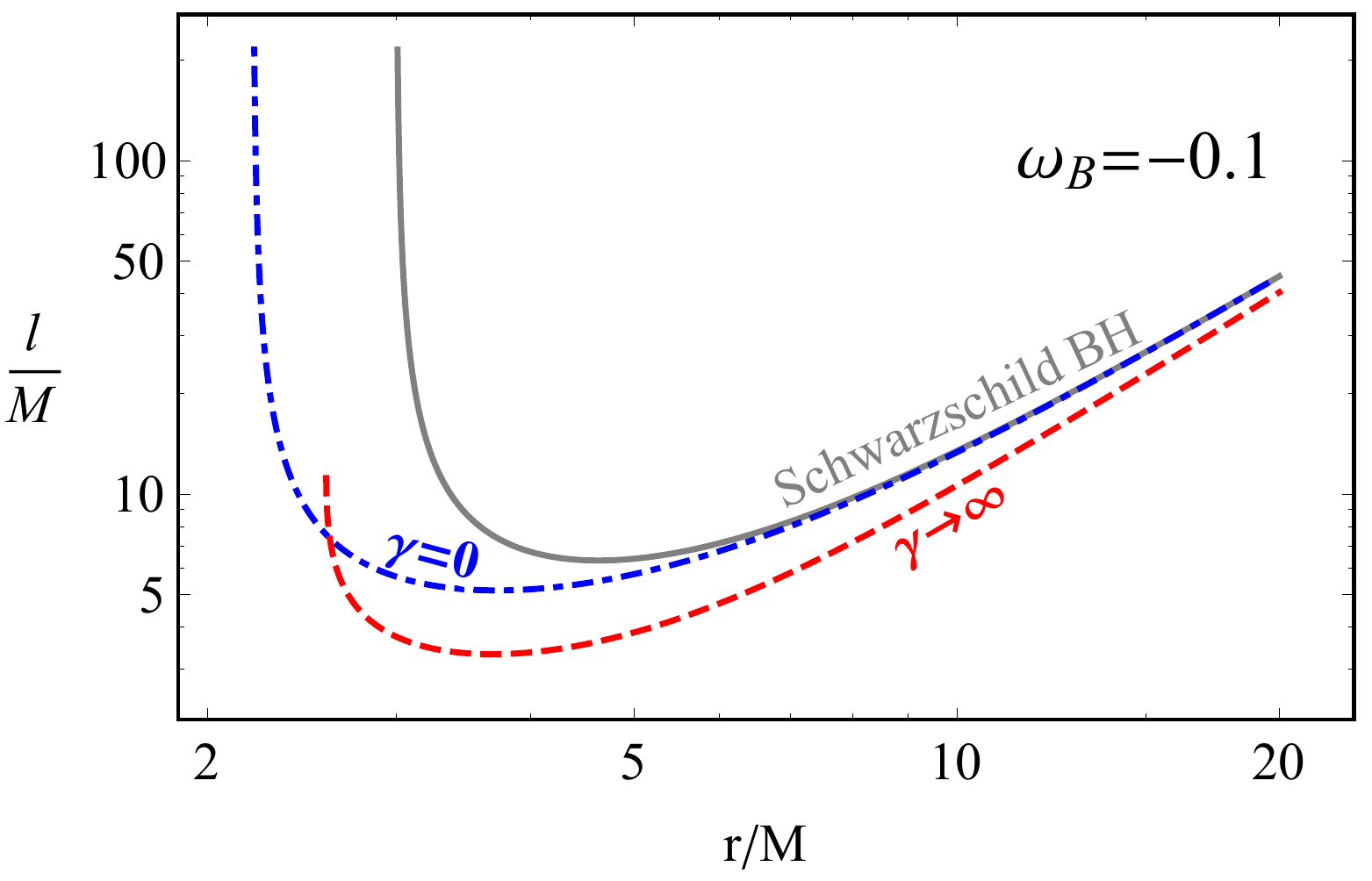}
	\caption{Dependence of specific angular momentum of positively (on the top panel) and negatively charged particles (at the bottom panel) for circular orbits from the radial coordinate for the different values of $\gamma$ at the fixed value of the parameter $\lambda=1$ with comparison to the Schwarzschild case. \label{lw}}
\end{figure}
 
 Figure~\ref{lw} demonstrates the specific angular momentum of charged particles with the values $\omega_{\rm B}=0.1$ and $\omega_{\rm B}=-0.1$ (in the top and bottom panels, respectively) for circular orbits for the different values of the parameters $\gamma$ fixing the parameter $\lambda=1$ with the comparison to the Schwarzschild case. One may see that the increase of the parameter $\gamma$ decreases the minimum distance for circular orbits for both positively and negatively charged particles with respect to the Schwarzschild case. The value of specific angular momentum decreases with increasing the $\lambda$ parameter and it increases with increasing $\gamma$ parameter.
 
 \begin{figure}[ht!]
   \centering
  \includegraphics[width=0.98\linewidth]{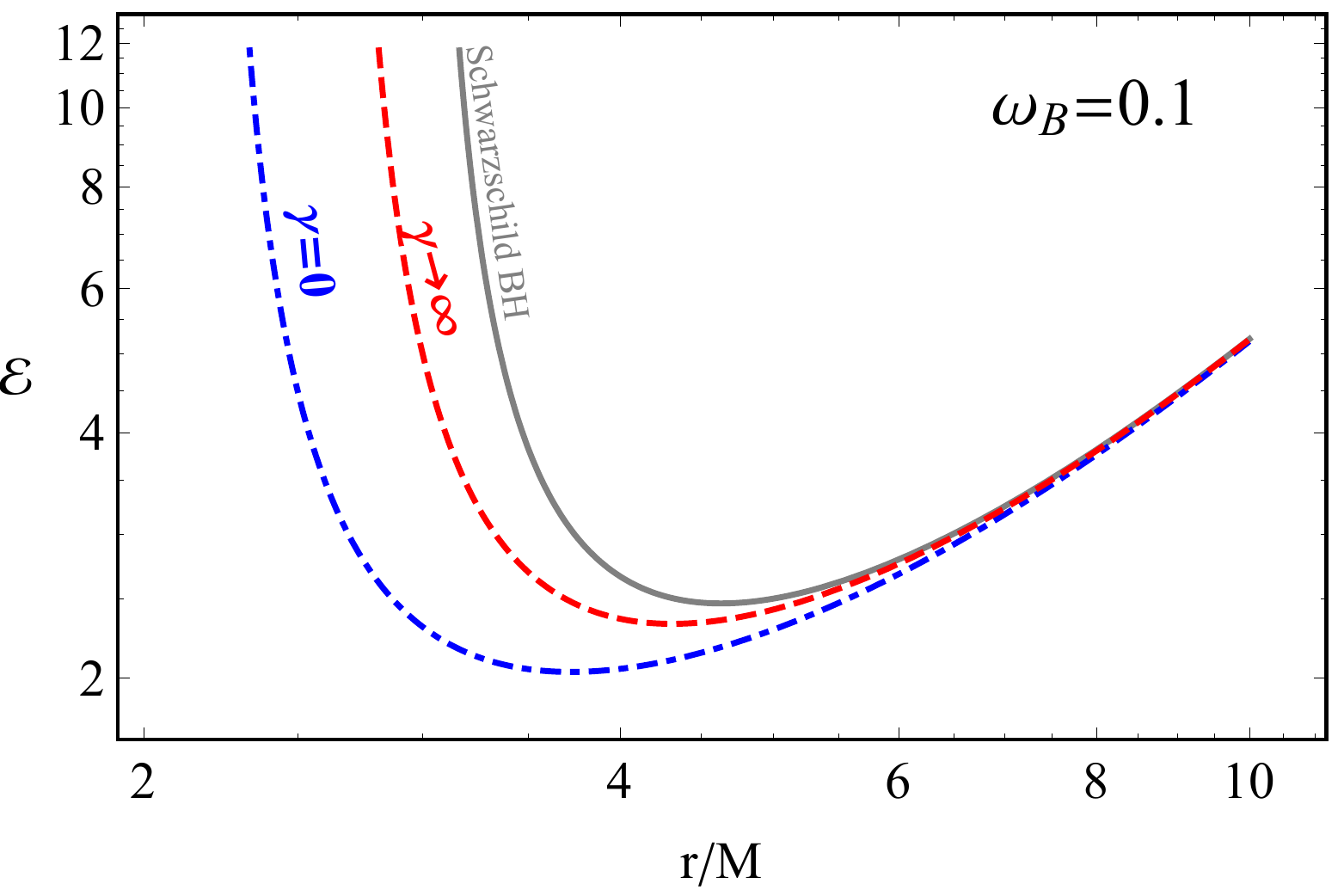}
  \includegraphics[width=0.98\linewidth]{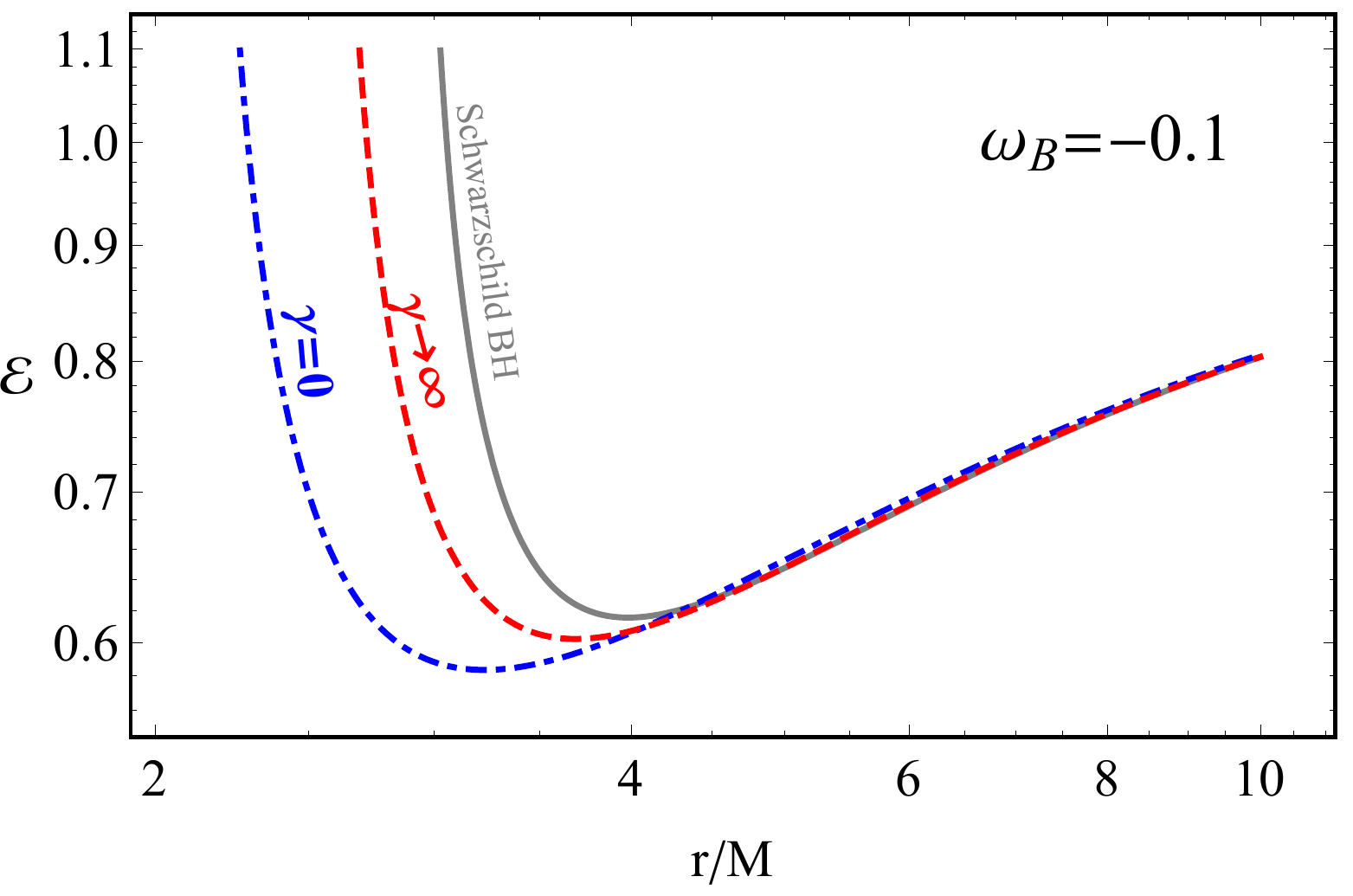}
	\caption{Dependence of specific energy of positively (on the top panel) and negatively charged particles (at the bottom panel) for circular orbits from the radial coordinate for the different values of $\gamma$ at the fixed value of the parameter $\lambda=1$ with comparison to the Schwarzschild case. \label{Ew}}
\end{figure}

Figure~\ref{Ew} illustrates the specific energy  of the charged particles with the values $\omega_{\rm}=0.1$ and $\omega_{\rm}=-0.1$ (in the top and bottom panels, respectively) which corresponds to circular orbits for the different values of the parameters $\gamma$ fixing the parameter $\lambda=1$ with comparison to the Schwarzschild case. One may see that the increase of the parameter $\gamma$ causes to decrease of the minimum distance for circular orbits and the distance where the specific energy minimum which corresponds to ISCO for both positively and negatively charged particles with respect to the Schwarzschild case. The value of specific energy decreases with increasing $\lambda$ parameter and it increases with increasing $\gamma$ parameter.  Moreover, one can see from comparison of panels in Figs.~\ref{lw} and \ref{Ew} that the positively charged particles can be in circular orbits having more specific energy and angular momentum with compare to that for the negatively charged particles due to the strong effect of the repulsive Lorentz force.

 One may derive the ISCO equation from the last part of the condition for stable circular orbits as following

 \begin{eqnarray} \label{iscoweq}
 \nonumber
 &&2 f(r) \Big\{2 r^2 \omega _B \left[\omega _B (2-2 r \mathcal{Z} (r \mathcal{Z}-1))+\mathcal{Q} \mathcal{Z} (3-2 r \mathcal{Z})\right]\\ \nonumber
 &&+\mathcal{Q}^2 (3-2 r \mathcal{Z})\Big\}+r^2 f''(r) \Big\{\Big[(r \mathcal{Z}-2)^2
\\
&&\pm 4 r \omega _B (r \mathcal{Z}-1) \left(\mathcal{Q} \pm r \omega _B (r \mathcal{Z}-1)\right)\Big]+\mathcal{Q}^2\Big\}\geq 0\ .
 \end{eqnarray}

ISCO radius of the charged particles can be described by Eq.(\ref{iscoweq}). However, it is impossible to solve it analytically and one can find the effects of the parameters $\gamma$,$\lambda$ and $\omega_{\rm B}$, by plotting the ISCO profiles solving Eq.(\ref{iscoweq}) numerically.

\begin{figure}[h!]
   \centering
  \includegraphics[width=0.98\linewidth]{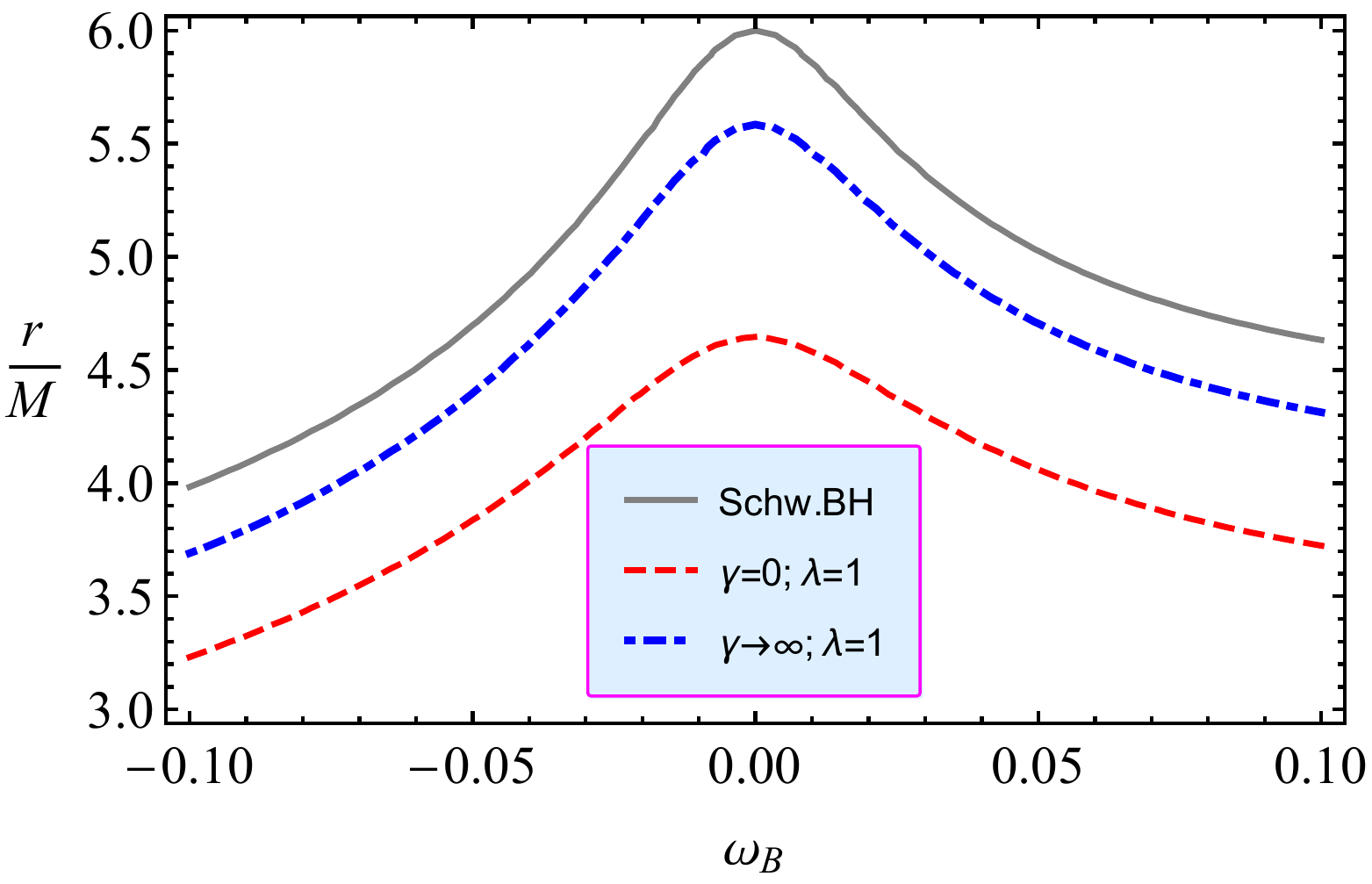}
	\caption{Dependence of ISCO radius from magnetic interaction parameter $\omega_{\rm B}$ for the different values of parameters $\lambda$ and $\gamma$, comparison with Schwarzschild BH. \label{iscow}}
\end{figure} 

Figure~\ref{iscow} demonstrates the dependence of ISCO radius from the magnetic interaction parameters $\omega_{\rm B}$ for the different values of the parameter $\gamma$ and $\lambda$. One may see that the increase of the parameters $\lambda$ and $\omega_{\rm B}$ cause to decrease ISCO radius while the effect of the parameter $\gamma$ increases it.  
\section{The motion of magnetized particles \label{magnetpartmotion}}

The magnetized particle motion may be considered as one of possible ways to test the gravity theory and corresponding spacetime metric around BH. Magnetized neutron stars observed as radio pulsars with the precisely measured periods of rotation can be approximated as test magnetized particle in the close enrironment of SrgA* in the center of Milky Way galaxy. No pulsar has been observed in the central parsec of SgrA* so far mainly due to the scattering of radio waves by dense, turbulent and ionized plasma in the SMBH close environment rather than by an intrinsic absence of radio pulsars in the vicinity to the supermassive black hole. However, the further search in the radio frequencies exceeding \~ 15 GHz would provide a real chance of detecting radio pulsars in the close SMBH environment. On the other hand recently observed hot spots in orbital motion near the event horizon of SrgA* can be also treated as magnetized test particles. In order to explore the properties of RGI Schwarzschild BHs in this section we aimed to describe the motion of magnetized particles in the presence of the external asymptotically uniform magnetic field around the BH using the following Hamilton-Jacobi equation~\cite{deFelice}
\begin{eqnarray}\label{H-J}
g^{\mu \nu}\frac{\partial {\cal S}}{\partial x^{\mu}} \frac{\partial {\cal S}}{\partial x^{\nu}}=-\Bigg(m-\frac{1}{2} {\cal D}^{\mu \nu}{\cal F}_{\mu \nu}\Bigg)^2\ ,
\end{eqnarray}
with the term characterizing  the interaction between the magnetized particle with non-zero magnetic moment and the external magnetic field expressed as ${\cal D}^{\mu \nu}{\cal F}_{\mu \nu}$. We will assume the dipolar structure of the magnetized particle's magnetic moment. The polarization tensor ${\cal D}^{\alpha \beta}$ must satisfy the following condition
\begin{eqnarray}\label{dexp}
{\cal D}^{\alpha \beta}=\eta^{\alpha \beta \sigma \nu}u_{\sigma}\mu_{\nu}\ , \qquad {\cal D}^{\alpha \beta }u_{\beta}=0\ ,
\end{eqnarray} 
where $\mu^{\nu}$ is the four-vector of dipole moment of the magnetized particle. It is possible to determine the interaction term of the Hamilton-Jacobi equation (\ref{HJ}) using the electromagnetic field tensor ${\cal F}_{\alpha \beta}$ which can be expressed in terms of electric $E_{\alpha}$ and magnetic $B^{\alpha}$ fields as
\begin{eqnarray}\label{fexp}
{\cal F}_{\alpha \beta}=u_{\alpha}E_{\beta}-u_{\beta}E_{\alpha}-\eta_{\alpha \beta \sigma \gamma}u^{\sigma}B^{\gamma}\ . 
\end{eqnarray}
Taking account the condition given in Eq.(\ref{dexp})  we have 
\begin{eqnarray}\label{DF1}
{\cal D}^{\alpha \beta}{\cal F}_{\alpha \beta}=2\mu^{\hat{\alpha}}B_{\hat{\alpha}}=2\mu B_0 {\cal L}[\lambda_{\hat{\alpha}}]\ ,
\end{eqnarray}
where $\mu =\sqrt{|\mu_{\hat{i}}\mu^{\hat{i}}|}$ is the norm of the magnetic dipole moment of the particle and ${\cal L}[\lambda_{\hat{\alpha}}]$ is a function of the effects of comoving frame of reference with the magnetized particle rotating around the BH being function of the space coordinates, as well as other parameters which defines the tetrads $\lambda_{\hat{\alpha}}$ attached from the fiducial comoving observer ~\cite{deFelice}.

As we mentioned above our aim is to study the effects of RGI Schwarzschild BHs on the magnetized particle and for simplicity we consider the magnetic interaction between the magnetized particle and external magnetic field is weak enough (due to ether weakness of the external magnetic field or the weakness of magnetized particle), so we can use the approximation of $\left({\cal D}^{\mu \nu}{\cal F}_{\mu \nu} \right)^2  \to 0$. Moreover, we also assume that the direction of the magnetic dipole moment of the magnetized particle to be perpendicular to the equatorial plane, with the components  $\mu^{i}=(0,\mu^{\theta},0)$. 

Since the external asymptotically uniform magnetic field does not break the spacetime symmetries we have still two conserved quantities of the motion magnetized particles: angular momentum $p_{\phi}= L$ and energy $p_t = -E$ of the particle, respectively. The expression for the action of magnetized particles with allows to separate variables in the Hamilton-Jacobi equation (\ref{H-J}) can be written as

\begin{eqnarray}\label{action}
{\cal S}=-E t+L\phi +{\cal S}_r(r)\ .
\end{eqnarray}

The expression for radial motion of magnetized particles at the equatorial plane where $\theta=\pi/2$ ($p_{\theta}=0$), substituting Eq.(\ref{DF1}) to Eq.(\ref{HJ}) taking account of Eq.(\ref{action}) can be found in the following form 
\begin{eqnarray}
\dot{r}^2={\cal{E}}^2-V_{\rm eff}(r;\gamma ,l,\beta)\ . 
\end{eqnarray}
The analytic expression of the effective potential takes the form
\begin{eqnarray}\label{effpot}
V_{\rm eff}(r;l,\beta,\gamma,\lambda)=f(r)\left(1+\frac{l^2}{r^2}-\beta {\cal L}[\lambda_{\hat{\alpha}}]\right)\ ,
\end{eqnarray}
where $\beta = 2\mu B_0/m$ is the magnetic coupling parameter which corresponds to the interaction term ${\cal D}^{\mu \nu}{\cal F}_{\mu \nu}$ in the Hamilton-Jacobi equation (\ref{HJ}). For the typical neutron star orbiting around supermassive black hole with magnetic dipole moment $\mu=(1/2)B_{\rm NS}R^3_{\rm NS}$ the magnetic parameter is 
\begin{eqnarray}\label{betaNS}
\beta \simeq 0.0044\left(\frac{B_{\rm NS }}{10^{12} \rm G}\right)\left(\frac{R_{\rm NS}}{10^6 \rm cm}\right)^3\left(\frac{B_{\rm ext}}{10\rm G}\right)\left(\frac{m_{\rm NS}}{M_{\odot}}\right)^{-1} \ . 
\end{eqnarray}
We recall the conditions for circular orbits
\begin{eqnarray} \label{conditions}
\dot{r}=0 \ , \qquad \partial_r V_{\rm eff}=0\ .
\end{eqnarray}
The first equation of (\ref{conditions}) with together Eq.~(\ref{effpot})
gives the following expression 
\begin{eqnarray}\label{betafunc1}
\beta(r;l,{\cal E},\gamma,\lambda)=\frac{1}{ {\cal L}[\lambda_{\hat{\alpha}}]}\Bigg(1+\frac{l^2}{r^2}-\frac{{\cal{E}}^2}{f(r)}\Bigg)\ .
\end{eqnarray}

Since we study the motion of the magnetized particle at the equatorial plane, we consider the components of the external magnetic field in that plane measured by the comoving observer,  which take the form
\begin{eqnarray}\label{Bcomp}
B_{\hat{r}}=B_{\hat{\phi}}=0\ , \qquad B_{\hat{\theta}}=B_0f(r)\,e^{\Psi}\ ,
\\{\rm with}\, \qquad \label{epsi}
 e^{-\Psi}=\left[f(r)-\Omega^2 r^2\right]^{\frac{1}{2}}\ ,
\end{eqnarray}
where 
\begin{eqnarray}
  \Omega=\frac{d\phi}{d t}=\frac{d\phi/d\tau}{d t/d\tau}=\frac{f(r)}{r^2}\frac{l}{{\cal{E}}}\ 
\end{eqnarray}
is the angular velocity of the magnetized particle.

The interaction term in Eq.~(\ref{H-J}) can be found using Eqs.(\ref{Bcomp}) and (\ref{DF1}) in the form
\begin{eqnarray}\label{DF2}
{\cal D}^{\mu \nu}{\cal F}_{\mu \nu}=2\mu B_0f(r)\,e^{\Psi}\ , \end{eqnarray}
which can be used to find the characteristic function 
\begin{eqnarray}\label{lambda}
{\cal L}[\lambda_{\hat{\alpha}}]=e^{\Psi}\, f(r)\ .
\end{eqnarray}
Finally, we will find the exact form of the magnetic coupling parameter $\beta(r;l,{\cal E},\alpha)$ using Eqs. (\ref{lambda}), (\ref{epsi}), and (\ref{betafunc1}) in the form 
\begin{eqnarray}\label{betafinal}
  \beta(r;l,{\cal E},\gamma,\lambda)=\left(\frac{1}{f(r)}-\frac{l^2}{{\cal E}^2 r^2}\right)^{\frac{1}{2}}\Bigg(1+\frac{l^2}{r^2}-\frac{{\cal E}^2}{f(r)}\Bigg).\ \
\end{eqnarray}

The Eq.(\ref{betafinal}) indicates that magnetized particle with given specific energy ${\cal E}$ and angular momentum $l$ can be at the circular orbit for a given radial distance $r$ from the center of BH.
 
\begin{figure}[ht!]
  \centering
   \includegraphics[width=1\linewidth]{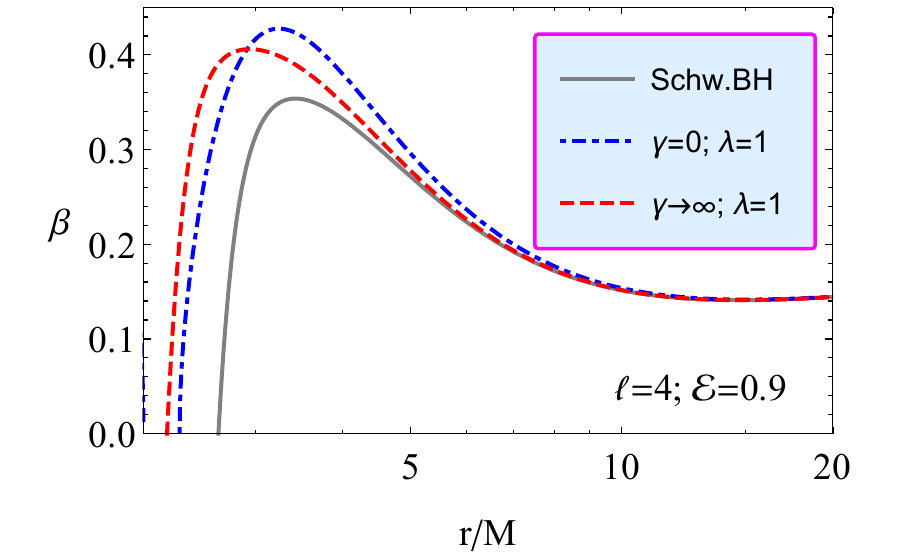}
   \caption{Radial profiles of the function of magnetic interaction parameter $\beta$ for the different values of the parameters $\gamma$ and $\beta$. For plotting the figures we have used the following values of the specific angular momentum $l=4$ and the energy of the particle as ${\cal E}=0.9$. \label{betafig}}  
\end{figure}
  
The magnetic coupling/interaction  parameter $\beta$ as a function of radial coordinate for the different values of $\beta$ parameter is demonstrated in Figure~\ref{betafig}. One may see that the increase of the parameter $\lambda$ ($\gamma$) causes to increase (decrease) of the maximum (minimum) value of the magnetic coupling parameter and the minimum distance where circular orbits for magnetized particle are allowed decreases with increasing both parameters $\gamma$ and $\lambda$.
  
Now the stable circular orbits of magnetized particles can be redefined in terms of $\beta$ by the following conditions
\begin{eqnarray}\label{conditionstab}
\beta =\beta(r;l,{\cal E},\gamma,\lambda), \qquad \partial_r \beta(r;l,{\cal E},\gamma,\lambda)=0\ .
\end{eqnarray}
The  solution of the system of equations with five variables $\beta,r,l,{\cal E}$ and the parameter $\gamma$ given in Eq.(\ref{conditionstab}) can be obtained in terms of any two of the five independent variables.

The expressions for minimum value for the specific energy ${\cal E}$ of a magnetized particle at stable circular orbit can be found  by solving the second part of Eq.~(\ref{conditionstab}) with respect to the energy ${\cal E}$ and using the coupling parameter $\beta$ together with radial coordinate $r$ as free parameters:
\begin{eqnarray}\label{emin}
{\cal E}_{\rm min}(r;l,\gamma,\lambda)=\frac{l \left[\lambda  M^2 \Omega (r+\gamma M) + r^2(r-2 M)\right]}{r^2 \sqrt{M r^3-\lambda  M^3 \Omega  (r+2 \gamma  M)}}\ .
\end{eqnarray}

In the limiting case when $\lambda \to 0$  the Eq.~(\ref{emin}) coincides with the corresponding relation for pure Schwarzschild BH  which has the form 
\begin{eqnarray}
{\cal E}_{\rm min}(r;l)=\frac{l}{\sqrt{Mr}}\left(1-\frac{2M}{r}\right)\ .
\end{eqnarray}

\begin{figure}[ht!]
  \centering
   \includegraphics[width=0.95\linewidth]{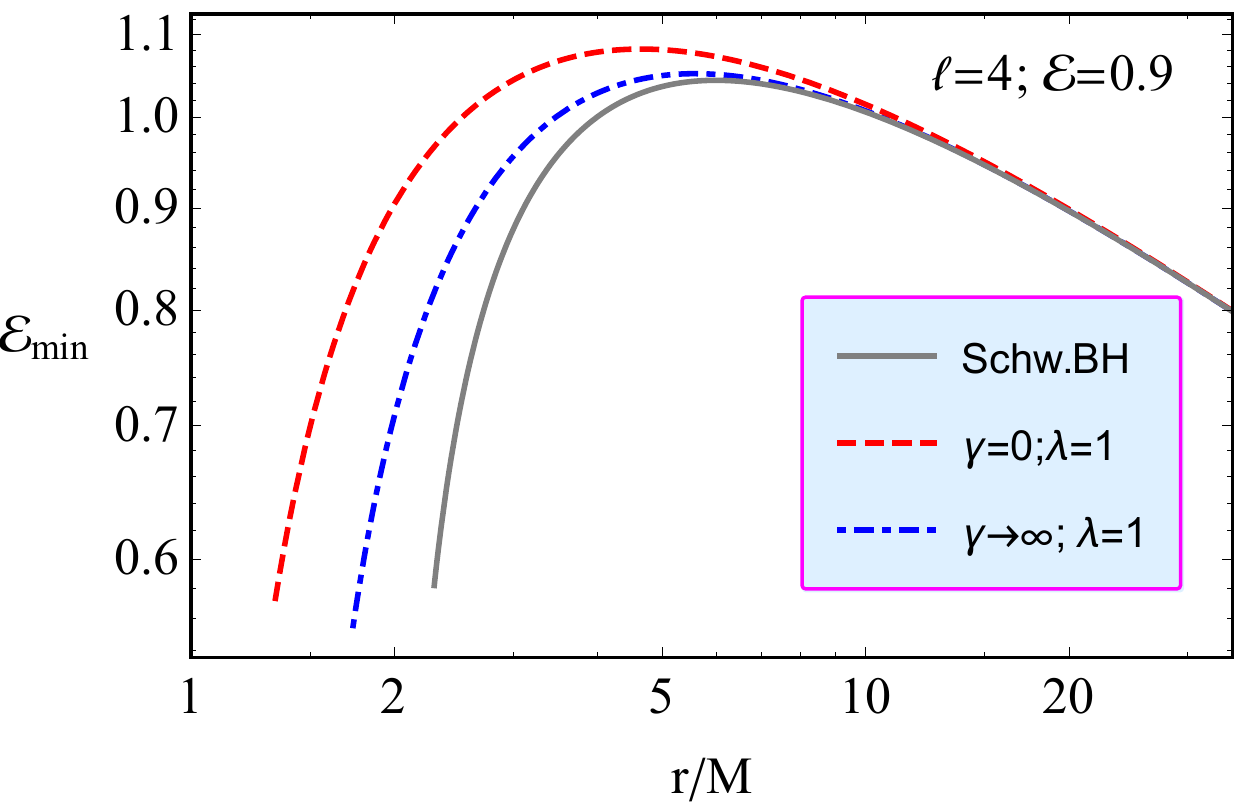}
      \caption{Radial profiles of the minimum specific energy of magnetized particles for the different values of the parameters $\gamma$ and $\lambda$, at $l=4$. \label{eminfig}}
\end{figure}

Figure~\ref{eminfig} presents the minimum value of specific energy of the magnetized particles allowing them to be in circular orbits as a function of the radial coordinate for the different values of the parameters $\gamma$ and $\lambda$. The increase of $\lambda$ parameter causes the increase of the  energy  near the central object while the increase of the parameter $\gamma$ causes the decrease of the energy. At large distances from the BH effects of both parameters almost disappear, which can be explained that at far distances the quantum effects do not play an important role. Moreover, the value of the specific energy at the minimum distance of circular orbits decreases with increasing the parameters $\lambda$ and $\gamma$.

The expression for minimum of magnetic interaction parameter $\beta$ can be obtained by substituting Eq.(\ref{emin}) in to Eq.(\ref{betafinal}) in the following form 
\begin{widetext}
\begin{eqnarray}\label{betamineq}
\nonumber
\beta_{\rm min}(r;l,\gamma,\lambda)&=&\frac{\sqrt{\lambda ^2 M^4 \Omega ^2 (\gamma  M+r)^2+\lambda  M^2 r^3 \Omega  [(2 \gamma -1) M+2 r]+r^5 (r-3 M)}}{\left(\gamma  \lambda  M^3 \Omega +\lambda  M^2 r \Omega -2 M r^2+r^3\right) \left(\lambda  M^3 r^4 \Omega  (2 \gamma  M+r)-M r^7\right)} \Big\{l^2 \Big[\lambda ^2 M^4 \Omega ^2 (\gamma  M+r)^2 
\\
&& +\lambda  M^2 r^3 \Omega  [(2 \gamma -1) M+2 r]+r^5 (r-3 M)\Big]+\lambda  M^3 r^4 \Omega  (2 \gamma  M+r)-M r^7\Big\}\ .
\end{eqnarray}
\end{widetext}

\begin{figure}[ht!]
  \centering
   \includegraphics[width=0.98\linewidth]{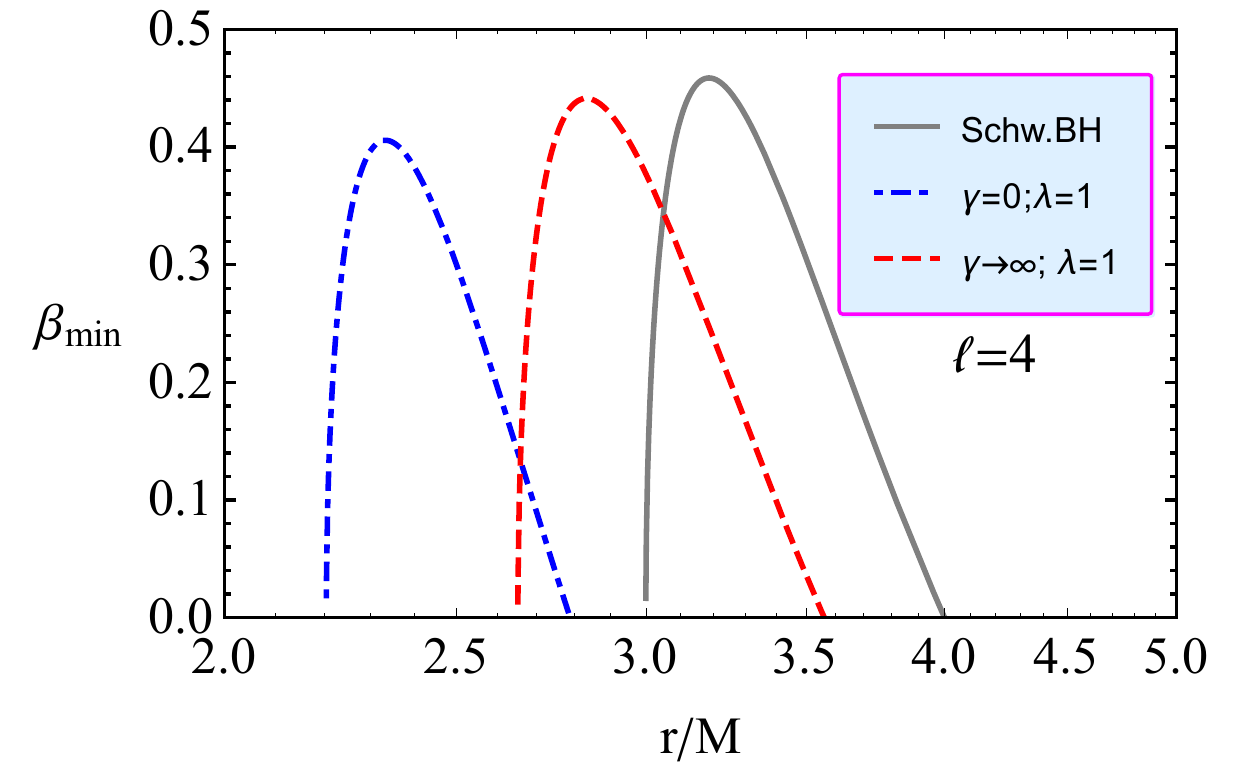}
    \caption{Radial profiles of minimum magnetic interaction parameter $\beta$ for the different values of the parameters $\gamma$ and $\lambda$. In plots we use the specific angular momentum as $l=4$. \label{betaminfig}}
\end{figure}

Figure~\ref{betaminfig} represents minimal magnetic interaction parameter as a function of radial coordinate for the different values of the parameters $\gamma$ and $\lambda$. One can see from the Fig.~\ref{betaminfig} that peak of the minimal magnetic coupling parameter decreases with the increase of the parameter $\lambda$ and the increase of the parameter $\gamma$ causes the increase of it.

The upper limit for stable circular orbits corresponds to minimum value of specific angular momentum. The extreme value of the magnetic interaction parameter corresponds to the minimum angular momentum and  can be found solving the equation $\partial_r \beta_{\rm min}(r;l,\gamma,\lambda)=0$ with respect to $l$:
\begin{widetext}
\begin{eqnarray}\label{lmineq}
\nonumber
l_{\rm min}(r; \gamma,\lambda)&=&\frac{r^3 \left(M r^3-\lambda  M^3 \Omega  (2 \gamma  M+r)\right)}{\sqrt{\lambda ^2 M^4 \Omega ^2 (\gamma  M+r)^2+\lambda  M^2 r^3 \Omega  [(2 \gamma -1) M+2 r]+r^5 (r-3 M)}}
\\
&&\left\{2 \lambda ^2 M^4 \Omega ^2 (r+\gamma  M)^2+\lambda  M^2 r^2 \Omega  \left[(4 \gamma -5) M r+4 r^2-6 \gamma  M^2\right]+r^5 (2 r-3 M)\right\}^{-1/2}\ .
\end{eqnarray}
\end{widetext}

\begin{figure}[h!]
  \centering
   \includegraphics[width=0.98\linewidth]{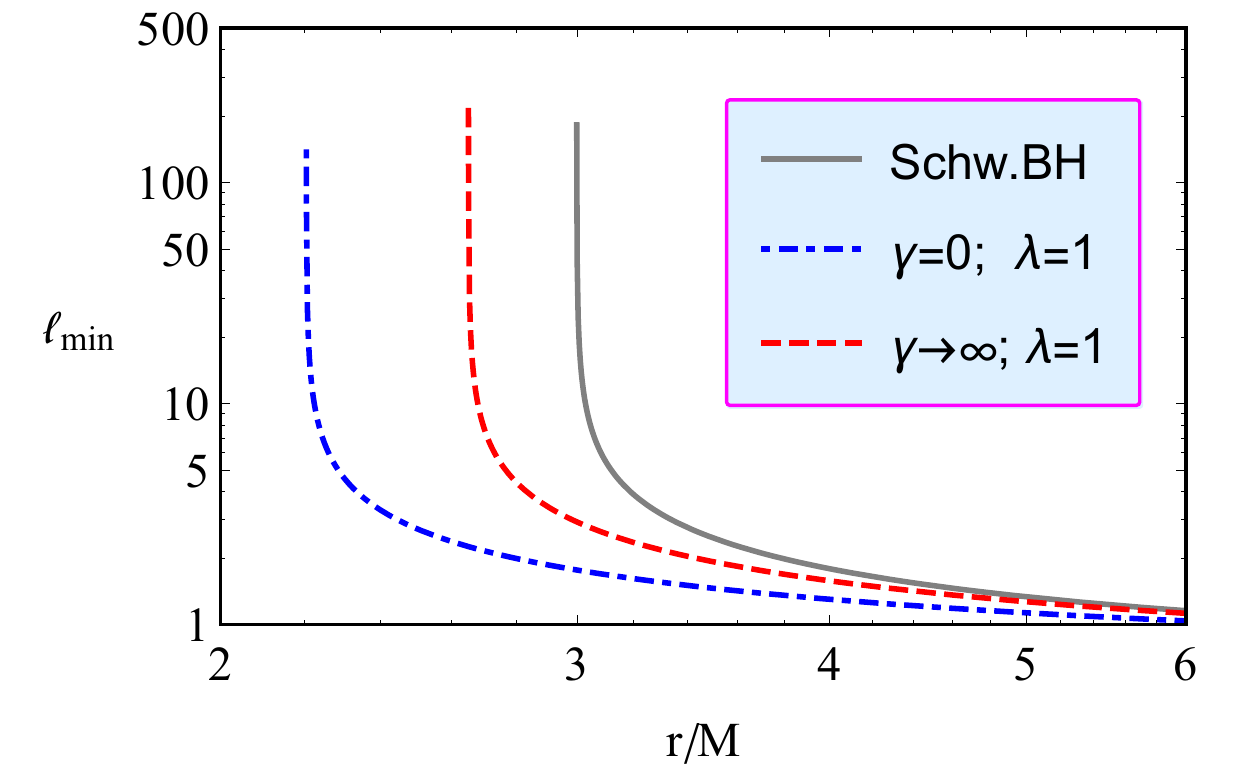}
    \caption{Radial profiles of minimal specific angular momentum, $l_{\rm min}$, for the different values of the parameter $\gamma$. \label{lminfig}}
\end{figure}

Figure \ref{lminfig} shows the minimum specific angular momentum as a function of radial coordinate for the different values of the parameter $\gamma$.  One may again see the similar effects of the RGI  parameters as it was just seen above: the increase of the parameter $\lambda$ causes the decrease of the minimum value for critic angular momentum  corresponding to circular stable orbits,  while the angular momentum increases with the increase of the parameter $\gamma$. 

One may find the extreme magnetic coupling parameter inserting Eq.(\ref{lmineq}) in to Eq.~(\ref{betamineq}) in the form: 
\begin{widetext}
\begin{eqnarray}
\beta_{\rm extr}(r;\gamma,\lambda)&=&\frac{4 \sqrt{\lambda ^2 M^4 \Omega ^2 (\gamma  M+r)^2+\lambda  M^2 r^3 \Omega  [(2 \gamma -1) M+2 r]+r^5 (r-3 M)}}{\lambda  M^2 \Omega  (\gamma  M+r)+r^3 }\nonumber \\ &\times &\left\{4-\frac{2 \lambda  M^3 r^2 \Omega  (6 \gamma  M+5 r)+6 M r^5}{\left[\lambda  M^2 \Omega  (\gamma  M+r)+r^3\right]^2}\right\}^{-1}
\end{eqnarray} \ .
\end{widetext}

\begin{figure}[h!]
  \centering
   \includegraphics[width=0.98\linewidth]{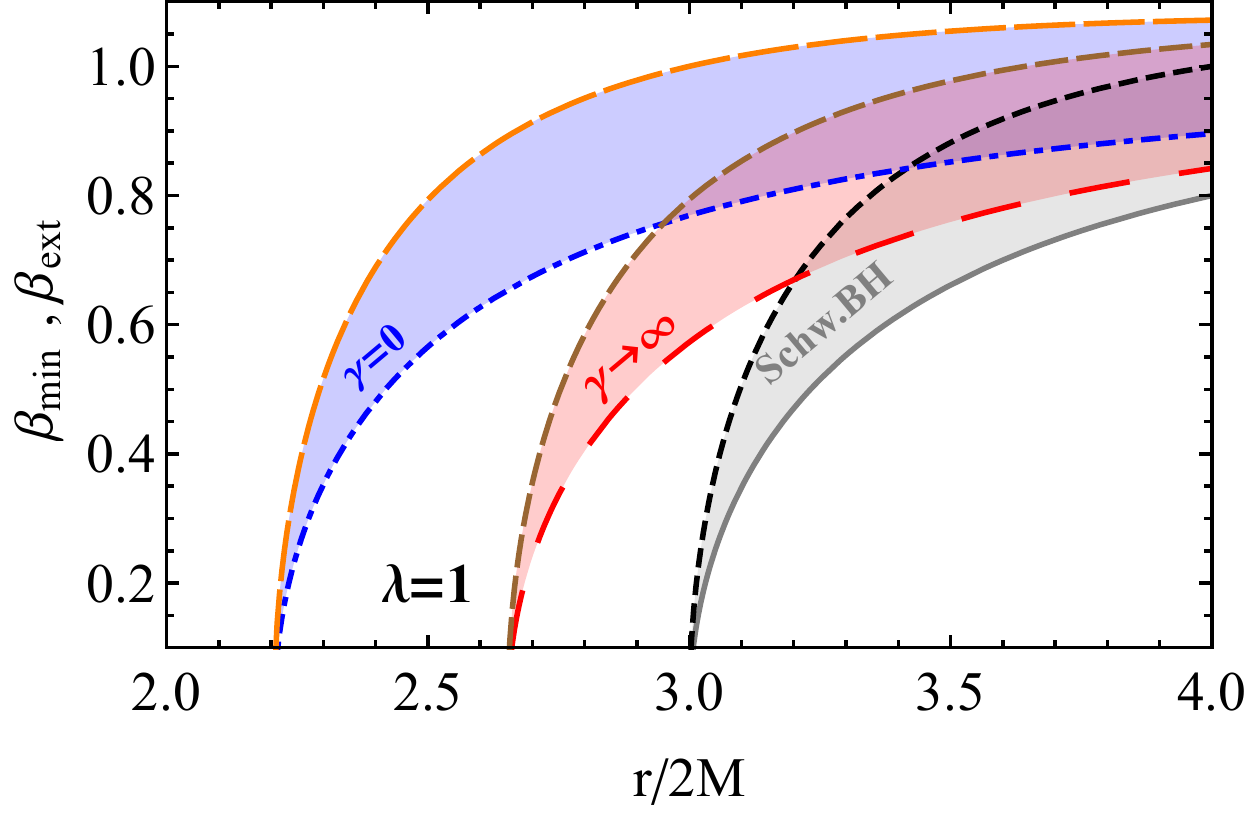}
    \caption{Radial profiles of the minimum magnetic coupling parameter of a magnetized particle with $l=0$ and extreme magnetic coupling parameter for the different values of the parameter $\lambda$ at $\gamma=0$ and $\gamma \to \infty$. Gray, light-blue and light-red colored areas correspond to the parameter $\lambda=0$ (Schwarzschild case), $\lambda=1, \  \gamma=0$ and $\lambda=1,\, \gamma \to \infty$, respectively.     \label{betaextremfig}}
\end{figure}

Figure \ref{betaextremfig} illustrates the radial dependence of the extreme value of the magnetic coupling parameter and the minimum value of the magnetic coupling parameter at $\beta_{\rm min}(l=0)$, for the different values of the parameter, $\gamma$. The colored areas imply the range where a magnetized particle with the magnetic coupling parameter $\beta_{\rm extr}<\beta<\beta_{\rm min}(l=0)$ stable circular orbits are allowed. The specific angular momentum of the particle at the range where circular orbits are allowed can be found solving the following equation

\begin{equation}
    \beta=\beta_{\rm min}(r; \lambda,\gamma)
\end{equation}
with respect to $l$ and we have
\begin{eqnarray}\label{lminbeta}
l_{\rm min}(r;\gamma ,\lambda;\beta)=\frac{r^2 \left(\beta -\sqrt{\frac{1}{f(r)}+\mathcal{C}(r;\gamma ,\lambda )}\right)}{[1-\mathcal{D}(r;\gamma ,\lambda )] \sqrt{\frac{1}{f(r)}+\mathcal{C}(r;\gamma ,\lambda )}}\ ,
\end{eqnarray}

One may find the minimum energy of the magnetized particle at the circular orbits  inserting Eq.~(\ref{lminbeta}) in to Eq.~(\ref{emin}) in the following form
\begin{eqnarray}
{\cal E}_{\rm min}(r;\gamma ,\lambda;\beta)=\frac{\mathcal{A}(r;\gamma ,\lambda ) \sqrt{\beta -\sqrt{\mathcal{C}(r;\gamma ,\lambda )+\frac{1}{f(r)}}}}{2 \sqrt[4]{\frac{1}{f(r)}+\mathcal{C}(r;\gamma ,\lambda )} \sqrt{r-\frac{\mathcal{D}(r;\gamma ,\lambda )}{r}}}\ ,
\end{eqnarray}
\begin{widetext}
where denotes
\begin{eqnarray}
&&\mathcal{C}(r;\gamma ,\lambda) =\frac{8 r^2 \left[2 \gamma  \sqrt{\gamma +2} (9 \gamma +2)^{3/2} \lambda -8 r^3+\sqrt{\gamma +2} (9 \gamma +2)^{3/2} \lambda  r-(9 \gamma  (3 \gamma +4)-4) \lambda  (2 \gamma +r)\right]}{\left[\gamma  \sqrt{\gamma +2} (9 \gamma +2)^{3/2} \lambda +8 (r-2) r^2+\sqrt{\gamma +2} (9 \gamma +2)^{3/2} \lambda  r-(9 \gamma  (3 \gamma +4)-4) \lambda  (\gamma +r)\right]^2}\ ,
\\
&&\mathcal{D}(r;\gamma ,\lambda )=\frac{ f(r,\gamma ,\lambda )}{\lambda  r^2 }\frac{\left[r^3+\frac{1}{8} \left(\sqrt{\gamma +2} (9 \gamma +2)^{3/2}-9 \gamma  (3 \gamma +4)+4\right) \lambda  (\gamma  +r)\right]^2}{r^3-\frac{1}{8} \left(\sqrt{\gamma +2} (9 \gamma +2)^{3/2}-9 \gamma  (3 \gamma +4)+4\right) (2 \gamma +r)}\ ,
\\
&&\mathcal{A}^2(r;\gamma ,\lambda) =8 r^3-2 \gamma  \sqrt{\gamma +2} (9 \gamma +2)^{3/2} \lambda -\lambda r\sqrt{\gamma +2} (9 \gamma +2)^{3/2} +(9 \gamma  (3 \left[ \gamma +4\right]-4) \lambda  (2 \gamma +r) \ .
\end{eqnarray}
\end{widetext}

In order to see the effect of the parameters $\gamma$ and $\lambda$ on the range where circular orbits are allowed we have carried out numerical calculations results of which are presented in Table~\ref{tab}. 

\begin{table}[h!] \begin{center}\caption{\label{tab} Numerical values for the circular orbits allowed area $\Delta r=r_{\rm max}-r_{\rm min}$ for the different values of the magnetic interaction parameter $\beta$ and the parameter $\gamma$.} \begin{tabular}{|c| c| c| c| c| c| c| }\hline
$ $ & $\beta=0.1$ & $\beta=0.1$ & $\beta=0.5$ & $\beta=0.5$ & $\beta=0.9$ & $\beta=0.9$ \\[1.ex]%\hline %
$ \lambda $ & $\gamma=0$ &  $\gamma \to \infty$ & $\gamma=0$  &  $\gamma \to \infty$ & $\gamma=0$ &  $\gamma \to \infty$ \\[1.ex]\hline %
$0.1 $ & 0.00415 & 0.00420 & 0.1336 & 0.1351 & 9.6857 & 9.7232 \\[1.5ex]\hline
$0.5 $ & 0.00396 &  0.00423 & 0.1284 & 0.1366& 9.3955 & 9.60921\\[1.5ex]\hline
$ 1 $ & 0.00383& 0.00441 & 0.1259 & 0.1429 & 8.9194 & 9.4511 \\[1.5ex] \hline
\end{tabular} \end{center}
 \end{table}

Table~\ref{tab} demonstrates the range of circular orbits of the magnetized particle around the RGI Schwarzschild BH different values of the magnetic interaction and the parameter $\gamma$.  One can see from the table~\ref{tab} that in the case when $\gamma=0$ the increase of the parameter $\lambda$ forces the range $\Delta r$ to shrink, while in the case when $\gamma \to \infty$ the area expands with the increase of parameter $\lambda$. 

\begin{eqnarray}
\lim_{r \to \infty }\beta_{\rm extr}=\lim_{r \to \infty }\beta_{\rm min}(l=0)=1
\end{eqnarray}

It implies the magnetized particle's stable circular orbits are placed at the infinity with $\beta=1$ in the other words the particle can not be in circular stable orbits with $\beta \geq 1$. For a typical neutron star orbiting around supermassive black hole SrgA* with 1.4 solar mass and radius 10 km, assuming average magnetic field in close environment of the supermassive black hole is about 100 G, calculations for the table orbits show that the neutron star's dipolar magnetic field at the surface should be less than 1.623$ \times 10^{13}$ G in order to be in circular orbit in the  close black hole environment. In the case of the magnetar detected on the Milky Way center  called SRG (PSR) J1745--2900 orbiting around Sgr A* ~\cite{Mori2013ApJ} at the distance 0.1 pc from SrgA*, with magnetic dipole moment $\mu \approx 1.6\times 10^{32} \rm \ G\cdot cm^3$ and mass $m \approx 1.5 M_{\odot}$, the magnetic coupling parameter is about 7.16. It implies that the magnetar orbits can not be stable in the close the supermassive black hole environment.

\section{Conclusion \label{conclusion}}

In this report we have studied in detail the dynamics of neutral, charged and magnetized particles around RGI Schwarzschild BH in the presence of external magnetic field and compared with the recent observational data on motion of the magnetar orbiting around the supermassive black hole SrgA* environment. The obtained results can be summarized as follows: 

\begin{itemize}

\item We have first analyzed the spacetime structure around RGI Scwarzschild BH through considering the Ricci and Kretschmann scalars. It was shown that the presence of both parameters of RGI Schwarzschild BH $\gamma$ and $\lambda$ cause the decrease of the values of the curvature scalars at the origin of the BH.
However when $\gamma=0$ the curvature scalars diverge at the center ans from the physical singularity. 

\item The analysis of the event horizon structure of RGI Schwarzschild BH has shown that with increasing the $\lambda$ parameter from 0 to 1 the distance between outer and inner horizons decreases and vanishes for $\lambda=1$. Beyond $\lambda>1$ we do not have any event horizon. 

\item We  studied  circular orbits of neutral particles around RGI Schwarzschild BH and showed that the energy (angular momentum) of the particles decreases (increases) with respect to pure Schwarzschild case with decreasing the photon sphere corresponding to marginally stable circular orbits.

\item The comparison the ISCO radius of particles around RGI and Kerr BHs provided an information that within the range of  $0<\gamma<\infty$ one may mimic the rotation parameter of Kerr BH up to $a/M \lesssim 0.31$. We also analyzed the energy efficiency of accretion disc around RGI BH and showed that efficiency increases with the increase of both $\lambda$ and $\gamma$ parameters.

\item We have studied the charged particle motion around RGI Schwarzschild BH in the presence of magnetic field. The analysis of circular orbits showed that both magnetic field and $\lambda$ parameter cause the decrease of the ISCO radius, while $\gamma$ parameter causes the increase of it. 

\item Using the Hamilton-Jacobi equation containing the electromagnetic interaction term between magnetic field and magnetic dipole moment of particle we have studied the magnetized particles motion in the backround test magnetic field approximation. It was shown that the minimum value of specific energy
of the magnetized particles increases with the increase of $\lambda$ and $\gamma$ parameters. 

\item It was also shown that the peak of the minimal magnetic coupling
parameter decreases with the increase of $\lambda$ parameter, while increase of $\gamma$ parameter causes the increase of $\beta_{\rm min}$. The range of circular orbits of the magnetized particle around the RGI Schwarzschild BH $\Delta r$ decreases with the increase of $\lambda$ while in the limit of when $\gamma=0$ the value of $\Delta r$ increases  with the increase of $\lambda$.

\item Moreover, the detailed analysis of stability of circular orbits shows that any magnetized particle can be in stable circular orbits when the coupling parameter $\beta \geq 1$ and it is shown that the orbit of magnetar of type SRG (PSR) J1745-2900 can not be  stable in the close vicinity of the supermassive black hole Sgr A* where the magnetic field strength is estimated as $B=100 \ \rm G$. In other words existence of the magnetars in the very close environment of SMBH is excluded by the stability analysis. 

\end{itemize} 
\section*{Acknowledgement}

This research is supported by Grants No. VA-FA-F-2-008, No.MRB-AN-2019-29 of the Uzbekistan Ministry for Innovative Development. JR AA and AB thank Silesian University in Opava for the hospitality during their visit. AA is supported by postdoc fellowship through the PIFI fund of Chinese Academy of Sciences. WH is supported by NSFC No. 11773059.

\bibliographystyle{apsrev4-1}
\bibliography{gravreferences}

\end{document}